\title{TransGameEnergyHarvesting_doublecolumn}

\documentclass[journal]{IEEEtran}
\ifCLASSINFOpdf
\else
\fi

\usepackage{times,epsfig,algorithmic,array}
\usepackage{mdwmath}
\usepackage{mdwtab}
\usepackage{eqparbox}
\usepackage{cite}
\usepackage{graphicx}
\usepackage{amsfonts}
\usepackage{epstopdf}
\usepackage{amsfonts}
\usepackage{amsmath}
\usepackage{amssymb}
\usepackage{amsthm}
\usepackage{algorithm}
\usepackage{algorithmic}

\begin{document}
%
\title{Transmission Game in MIMO Interference Channels With Radio-Frequency Energy Harvesting}
%
%
%

\author{Liang~Dong,~\IEEEmembership{Senior Member,~IEEE}
\thanks{L.~Dong is with the Department
of Electrical and Computer Engineering, Baylor University, Waco,
TX 76798, USA (e-mail: liang\_dong@baylor.edu).}
}

\maketitle

\begin{abstract}
For multi-user transmissions over MIMO interference channels, each user designs the transmit covariance matrix to maximize its information rate.  When passive radio-frequency (RF) energy harvesters are present in the network, the transmissions are constrained by both the transmit power limits and the energy harvesting requirements.  A passive RF energy harvester collects the radiated energy from nearby wireless information transmitters instead of using a dedicated wireless power source.  It needs multiple transmitters to concentrate their RF radiation on it because typical electric field strengths are weak.   In this paper, strategic games are proposed for the multi-user transmissions.  First, in a non-cooperative game, each transmitter has a best-response strategy for the transmit covariance matrix that follows a multi-level water-filling solution.  A pure-strategy Nash equilibrium exists.  Secondly, in a cooperative game, there is no need to estimate the proportion of the harvested energy from each transmitter.  Rather, the transmitters bargain over the unit-reward of the energy contribution.  An approximation of the information rate is used in constructing the individual utility such that the problem of network utility maximization can be decomposed and the bargaining process can be implemented distributively.  The bargaining solution gives a point of rates that is superior to the Nash equilibria and close to the Pareto front.  Simulation results verify the algorithms that provide good communication performance while satisfying the RF energy-harvesting requirements.


\end{abstract}



%
\IEEEpeerreviewmaketitle

\section{Introduction}

Energy harvesting delivers the necessary energy to operate the untethered IoT devices and alleviate the complete dependence on batteries~\cite{energy_scavenging,battery_free_wireless,4300988,5475111,7120024}.  Energy harvesting transducers can harvest energy from ambient sources and convert it to useful electrical energy.  Radio-frequency (RF) inductive devices convert electromagnetic energy to electrical energy.   
RF energy source is relatively omnipresent compared with solar, thermal, and vibration energy sources.  RF energy harvesting can be used in tandem with other energy harvesting technologies.  For example, the scavenged RF energy can prevent a solar-charged battery from discharging at night~\cite{energy_harvesting_ashrae}.   The harvester output can be in the forms of direct power or stored energy that charges super-capacitors or lithium ion batteries.  In particular, passive RF energy harvesting is promising that it exploits ambient RF sources instead of intentional high-power RF sources~\cite{7470500,7555321,dong_modeling_UHF}.  

However, RF energy harvesting has a drawback. Typical electric field strengths are weak (unless located close to sources), which severely limits the quantity of energy that can be harvested.  The passive RF energy harvesting of today's technology can only generate energy approximately an order of magnitude less than indoor solar and thermoelectric devices.  
To overcome this, we will coordinate multiple wireless transmissions of nearby communication systems and purposely concentrate the RF radiation at the energy harvesters.  Instead of working on the energy harvester side, we develop transmission techniques on the wireless information systems side to simplify the design of the IoT devices.   

In wireless communications, multi-user transmissions in interference channels imply competition and cooperation.  If the transmissions are over multiple-input multiple-output (MIMO) channels, each user designs its signal transmit covariance matrix to maximize the information rate while inevitably generating inference to others.  When there are RF energy harvesters in the network, the multi-user transmission problem needs to take into account both the communication constraints and the energy-harvesting requirements.   As multiple antennas are used at the communication transceivers, the energy is beamformed to the harvesters.   Coordinating multiple wireless transmitters for RF concentration is less demanding than interference management among the wireless communication systems.  We just need to ``accumulate'' the interference power at the energy harvesters.  

Simultaneous wireless information and energy transfer has attracted significant attention recently~\cite{7120025,7154495,7471533,7744827}.   The fundamental performance limits and the optimal transmission strategies have been studied in various communication scenarios such as the downlink of a cellular system~\cite{6589954}, the cooperative relay system~\cite{6552840}, and the broadcasting system~\cite{6489506,6373669}.  Also, several transmission strategies and power allocation methods have been proposed for multi-user multiple-input single-output (MISO) scenario~\cite{6860253,6760603}.  For a MIMO wireless broadcast system where one receiver harvests energy and other receivers decode information, transmission strategy is designed for maximal information rate versus energy transfer~\cite{6489506}.  With multiple single-antenna transmitters that send independent messages to their respective receivers, collaborative energy beamforming is required for wireless power transfer~\cite{6894630}.  Furthermore, there have been several studies of joint wireless information and energy transfer in interference channels~\cite{6253063,6571308,6697937,6893021,6861455,7008497}.  Because the interference has different impacts on the performances of information processing and energy harvesting at the receivers, the transmission strategy is a critical issue.  In a $K$-user MIMO interference channel, the transmitters transferring energy exploit a rank-one energy beamforming, and an iterative algorithm can be used to optimize the transmit covariance matrices~\cite{6861455}.  In another $K$-user MIMO scenario, the total transmit power of the users is minimized by jointly designing transmit beamformers, power splitters, and receive filters~\cite{7230306}. 
When a multi-antenna transmitter communicates with its information receiver, it can intentionally focus the transmitted power to the nearby RF energy harvesters with a guarantee on its information rate~\cite{Xing2017}.  

In this paper, a multi-user transmission problem is considered in MIMO interference channels with RF energy harvesting requirements.  
Each transmitter can estimate its user channel and measure the combined interference and noise.  However, the signaling between different users is very limited so that collaborative beamforming or joint design of transmit beamformers is not possible.  Instead, the transmitters compete by designing signal transmit covariance matrices to increase own information rates, subject to the transmit power constraints and the energy harvesting requirements.   Each transmitter knows the channel to the energy harvesters, however, it does not know the channels from other transmitters to the energy harvesters.

The multi-user transmission problem is formulated as a strategic game.  In a non-cooperative game, with a measurement of the combined interference and noise, the best response of a transmitter is to design the transmit covariance matrix according to the multi-level water-filling solution.  A pure-strategy Nash equilibrium exists.  However, it may not be unique or Pareto-efficient.  The best-response dynamic may cycle and not converge to a Nash equilibrium.   In a cooperative game, the sum information rate is maximized, and the optimal point of rates can be found that approaches the Pareto front~\cite{4604732,6193128}.  There is no need to estimate the proportion of the harvested energy from each transmitter.  Rather, the transmitters bargain over the unit-reward of the energy contribution to the harvesters. The individual utility function of the game is constructed from an approximation of the information rate such that the problem of network utility maximization can be decomposed and the bargaining process implemented distributively.    With moderate signaling among the users, the bargaining process quickly generates a point of rates that is close to the Pareto front.

The remainder of this paper is organized as follows.  In Section~\ref{sec:model}, a system model is provided for the multi-user MIMO communications, and the rate maximization problem is formulated with the RF energy harvesting requirement.   In Section~\ref{sec:noncooperative_game}, a non-cooperative game is proposed for the multi-user transmissions over interference channels.  The best-response strategy is based on a multi-level water-filling solution.
In Section~\ref{sec:cooperativegame}, a cooperative game is proposed for the multi-user transmissions.  The transmitters bargain over the unit-reward of the energy contribution while maximizing the network information rate.  
In Section~\ref{sec:numericalresults}, simulation results verify the transmission algorithms and show that the distributed bargaining offers superior performance.  Finally, conclusions are drawn in Section~\ref{sec:conclusion}.

\section{System Model and Problem Formulation}
\label{sec:model}

\begin{figure}[!t]
\centering
\includegraphics[width=5.1cm]{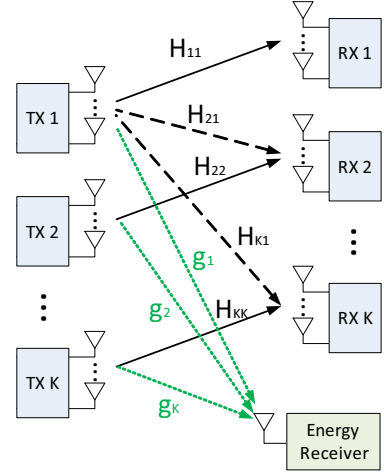}
\caption{$K$ pairs of wireless communication transmitter and receiver as well as an energy receiver.}
\label{fig:systemmodel}
\end{figure}

In the vicinity of the passive RF energy harvester, there are $K$ pairs of wireless communication transmitter and receiver that share the same radio channel resources of time and frequency (Figure~\ref{fig:systemmodel}).  Each transmitter has $M_{t}$ antennas and each receiver has $M_{r}$ antennas.   The passive RF energy harvester has a single antenna through which it can harvest the energy for the attached IoT device or transmit through backscattering~\cite{194920}.   In Sections~\ref{sec:model} through \ref{sec:distributedbargaining}, we develop the wireless transmission technique considering only one RF energy harvester.  The transmission methods will be extended to the scenario of multiple RF energy harvesters in Section~\ref{sec:multipleharvesters}.

Suppose that the wireless transmissions of the communication systems are narrowband over quasi-static fading channels.  In the baseband equivalent model, the signal transmitted at transmitter $i$ is $\mathbf{x}_{i} \in \mathbb{C}^{M_{t}}, i \in \mathcal{K} = \{1,2,\ldots, K\}$. 
The signal received at receiver $i$ is $\mathbf{y}_{i} \in \mathbb{C}^{M_{r}}, i \in \mathcal{K}$, which is given by
\begin{equation}\label{eq:baseband_receive}
\mathbf{y}_{i} = \mathbf{H}_{ii}\mathbf{x}_{i} + \sum_{j \in \mathcal{K} \setminus \{i\}} \mathbf{H}_{ij}\mathbf{x}_{j} + \mathbf{z}_{i}
\end{equation}
where $\mathbf{H}_{ii} \in \mathbb{C}^{M_{r} \times M_{t}}$ is the normalized user channel from  transmitter $i$ to its intended receiver, $\mathbf{H}_{ij} \in \mathbb{C}^{M_{r} \times M_{t}}$ is the normalized interference channel from transmitter $j$ to receiver $i$, and $\mathbf{z}_{i} \in \mathbb{C}^{M_{r} \times 1}$ is a zero-mean circularly symmetric complex Gaussian noise vector with $\mathbf{z}_{i} \sim \mathcal{CN}(\mathbf{0},\sigma_{n}^{2}\mathbf{I})$.  It is assumed that $\mathbf{z}_{i}$ contains both noise and possible interference generated by the backscatter device.  In this paper, without loss of generality, it is assumed that $\sigma_{n}^{2} =1$ for clarity.

The energy harvester (or the RF back scatterer) does not need to convert the received signal from the RF band to the baseband.  Nevertheless, the RF power is proportional to the power of the baseband signal.  Therefore, we use the baseband equivalent model to reveal the RF energy harvesting effects as well.   The baseband received signal at the energy harvester is given by
\begin{equation}
u = \sum_{i \in \mathcal{K}} \mathbf{g}_{i}^{H}\mathbf{x}_{i} +  n
\end{equation}
where $\mathbf{g}_{i} \in \mathbb{C}^{M_{t} \times 1}$ is the conjugate channel vector from transmitter $i$ to the energy harvester, and $n$ is the background noise.  

Let $\mathbf{Q}_{i}$ denote the covariance matrix of signal $\mathbf{x}_{i}$,  i.e., $\mathbf{Q}_{i} = \mathrm{E}[\mathbf{x}_{i} \mathbf{x}_{i}^{H}]$.  It is assumed that the Gaussian codebook with infinitely many codewords is used for the symbols and the expectation is taken over the entire codebook.  Therefore, $\mathbf{x}_{i}$ is zero-mean circularly symmetric complex Gaussian with $\mathbf{x}_{i} \sim \mathcal{CN}(\mathbf{0},\mathbf{Q}_{i})$.  The covariance matrix is Hermitian positive semidefinite that is denoted as $\mathbf{Q}_{i} \succeq0$.
The transmit power at transmitter $i$ is limited by its power constraint $P_{i}$, i.e., $\mathrm{Tr}(\mathbf{Q}_{i}) \leq P_{i}$.  Let $\mathbf{Q}_{-i}$ denote the collection of the transmit covariance matrices of the $(K-1)$ transmitters other than transmitter $i$.
From an information-theoretical perspective~\cite{Infromation_Cover,5247037}, the maximum information rate of transmitter-receiver pair $i$ is given by
\begin{equation}
r_{i}(\mathbf{Q}_{i}, \mathbf{Q}_{-i}) = \log \left|\mathbf{I} + \mathbf{R}_{i}^{-1} \mathbf{H}_{ii} \mathbf{Q}_{i}\mathbf{H}_{ii}^{H} \right|
\end{equation}
where $\mathbf{R}_{i}$ is the multi-user interference and noise as
\begin{equation}
\mathbf{R}_{i} = \sum_{j \in \mathcal{K} \setminus \{i\}} \mathbf{H}_{ij}\mathbf{Q}_{j}\mathbf{H}_{ij}^H + \mathbf{I}. \label{eq:information_rate}
\end{equation}

The received power at the RF energy harvester is the harvested energy normalized by the baseband symbol period.  It can be written as
\begin{equation}{\label{eq:harvestedpower}}
\zeta \mathrm{E}[|u|^{2}] = \zeta \sum_{i \in \mathcal{K}} \mathbf{g}_{i}^{H} \mathbf{Q}_{i} \mathbf{g}_{i}
\end{equation}
where $\zeta$ is a constant that indicates energy conversion efficiency.  
It is implied in \eqref{eq:harvestedpower} that, at the energy harvester, the noise power is negligible compared with the received signal power.  This makes sense because we do not consider any energy that can be harvested from noise.

For each communication receiver, the signals transmitted from other transmitters become the interference.   The received signals from other transmitters should be minimized.  In contrast, for the energy harvester, all of the transmitters are energy contributors.  The combined received signal power should be maximized.  At transmitter $i, i \in \mathcal{K}$, the design objective for the transmit covariance matrix $\mathbf{Q}_{i}$ is to maximize the achievable information rate while satisfying the individual transmit power constraint and the overall energy harvesting requirement.  The optimization problem can be formulated as ($\forall i \in \mathcal{K}$)
\begin{equation}\label{eq:problem1}
\mathcal{P}_{1}: \begin{array}{ll}
\begin{split}
\mathop{\text{maximize}}_{\substack{\mathbf{Q}_i}} 
\end{split}  
& r_{i}(\mathbf{Q}_{i},\mathbf{Q}_{-i})\\
\mathrm{subject~to} & \mathbf{Q}_{i} \succeq \mathbf{0} \\
& \mathrm{Tr}(\mathbf{Q}_i) \leq P_{i} \\
& \sum_{k \in \mathcal{K}} \mathbf{g}_{k}^{H} \mathbf{Q}_{k} \mathbf{g}_{k} \geq \Gamma 
\end{array}
\end{equation}
where $\Gamma$ indicates the energy harvesting requirement such that $\zeta \Gamma$ is the minimum required harvested energy during a symbol period. 
Because of the objective function and the last constraint of Problem $\mathcal{P}_{1}$, the design of transmit covariance matrix $\mathbf{Q}_{i}$ is coupled with those of other transmitters.  

\section{Wireless Transmissions in a Non-cooperative Game}
\label{sec:noncooperative_game}

\subsection{Non-cooperative Game}

Of the $K$ wireless communication users in the network, each user designs the transmit signal by regulating its covariance matrix to boost the maximum information rate.  Each user's transmission is within the transmit power limit and the overall radiated energy reaching the RF energy harvester should meet the energy harvesting requirement.  For each transmitter, modifying the transmit covariance matrix affects the interference to other receivers hence changing the maximum information rates of other users.  It is a competition among the wireless users; each tends its own maximum information rate.  Therefore, the problem of simultaneous wireless transmissions can be formulated as a strategic game~\cite{AlgorithmicGameTheory}, which is denoted as
\begin{equation}
\Game: \langle \mathcal{K}, \{S_{i}\}_{i \in \mathcal{K}}, \{u_{i}\}_{i \in \mathcal{K}} \rangle
\end{equation}
where $\mathcal{K}$ is the finite set of the $K$ transmitters as the game players, $S_{i}$ ($i \in \mathcal{K}$) is the set of strategies available to player $i$.  The overall strategy set of all the players is $S = \times_{i \in \mathcal{K}} S_{i}$.  The utility function of player $i$ ($i \in \mathcal{K}$) is  $u_{i}: S \rightarrow \mathbb{R}$. The transmit covariance matrix $\mathbf{Q}_{i}$ is the game strategy of player $i$, i.e., $\mathbf{Q}_{i} \in S_{i}$.   
In this paper, we address the general case in which each transmitter can freely adjust its transmit covariance matrix.  For some special cases with space-time coding or confined transmission, other strategies may be adopted.  For example, if the transmission is confined to the eigenmodes of the channel matrix, each transmitter can adopt the power allocation on the channel eigenmodes as the game strategy.  The maximum information rate of player $i$ is used as its utility function, i.e., $u_{i} = r_{i}(\mathbf{Q}_i,\mathbf{Q}_{-i})$.

With all competition but no cooperation among the players, this strategic game is a non-cooperative game.  
A steady state of the non-cooperative game is the Nash equilibrium (NE), where no player can improve his utility by unilaterally changing his own strategy.  In Game $\Game$, the NE can be expressed as
\begin{equation}
r_{i}(\mathbf{Q}_{i}^{*}, \mathbf{Q}_{-i}^{*}) \geq r_{i}(\mathbf{Q}_{i}, \mathbf{Q}_{-i}^{*}), \,\,\, \forall i \in \mathcal{K}
\end{equation}
where $\mathbf{Q}_{i}^{*}$ is the transmit covariance matrix for user $i$ at the NE and $\mathbf{Q}_{-i}^{*}$ is the collection of the transmit covariance matrices of all of the other users except user $i$ at the NE.

When $\mathbf{Q}_{-i}$ is fixed in Problem $\mathcal{P}_{1}$, the objective function is concave and the constraints are convex in the space of  positive semidefinite matrices.  Therefore, Problem $\mathcal{P}_{1}$ is a convex optimization problem for which efficient numerical optimization is possible~\cite{ConvexOpt_Boyd}.  In Game $\Game$, every strategy set $S_{i}$ is compact and convex, the utility function $u_{i} = r_{i}(\mathbf{Q}_{i}, \mathbf{Q}_{-i})$ is a continuous function in the profile of strategies $\mathbf{Q} \in S$ and concave in $\mathbf{Q}_{i}$.  Therefore, the game has at least one pure-strategy NE~\cite{10.2307/2032478,GameTheoryWirelessComm}.  
The NE is a strategy profile for which each player's strategy is a best response to the other players' strategies.  Given $\mathbf{Q}_{-i}$, transmitter $i$ optimizes the transmit covariance matrix $\mathbf{Q}_{i}$ as its best response by solving the following optimization problem
\begin{equation}\label{eq:problem_noncooperative}
\begin{array}{ll}
\begin{split}
\mathop{\text{maximize}}_{\substack{\mathbf{Q}_i}} 
\end{split}  
& \log\left| \mathbf{I} + \mathbf{R}_{i}^{-1} \mathbf{H}_{ii} \mathbf{Q}_{i} \mathbf{H}_{ii}^{H} \right|\\
\mathrm{subject~to} & \mathbf{Q}_{i} \succeq \mathbf{0} \\
& \mathrm{Tr}(\mathbf{Q}_i) \leq P_{i} \\
& \mathbf{g}_{i}^{H} \mathbf{Q}_{i} \mathbf{g}_{i} \geq \Gamma  - \sum_{k \in \mathcal{K} \setminus \{i\} } \mathbf{g}_{k}^{H} \mathbf{Q}_{k} \mathbf{g}_{k} = \Gamma_{i}
\end{array}
\end{equation}
where $\Gamma_{i}$ is a portion of the energy harvesting requirement imposed on transmitter $i$.  It can be acquired, for example, by shutting off user $i$'s transmission alone.  Suppose that the RF energy harvester can measure the harvested power (or energy over a symbol period) and feed it back to the transmitters through the Internet.  As the energy harvester is attached to an IoT device, this is feasible.  Transmitter $i$ turns on and off its transmission and observes the difference of the harvested energy from the feedback.  If there is mechanism to ensure that it is unlikely that multiple transmitters shut off at the same time, transmitter $i$ can estimate the minimum energy that it should contribute to the RF energy harvester.  

\subsection{Local Optimization Problem}

Transmitter $i$ can estimate its user channel $\mathbf{H}_{ii}$ through pilot signals and feedback, and it can measure the multi-user interference and noise $\mathbf{R}_{i}$ without the need to acquire individual $\mathbf{Q}_{j}$ or $\mathbf{H}_{ij}, j \neq i$. The objective function of the optimization problem \eqref{eq:problem_noncooperative} is
\begin{equation}
\log\left| \mathbf{I} + \mathbf{R}_{i}^{-1} \mathbf{H}_{ii} \mathbf{Q}_{i} \mathbf{H}_{ii}^{H} \right|
= \log\left| \mathbf{I} + \hat{\mathbf{H}}_{ii} \mathbf{Q}_{i} \hat{\mathbf{H}}_{ii}^{H} \right|
\end{equation}
where $\hat{\mathbf{H}}_{ii} = \mathbf{R}_{i}^{-1/2} \mathbf{H}_{ii}$.  A singular value decomposition on $\hat{\mathbf{H}}_{ii}$ gives $\hat{\mathbf{H}}_{ii} = \mathbf{U}_{i} \mathbf{\Sigma}_{i} \mathbf{V}_{i}^{H}$.  With $\hat{\mathbf{Q}}_{i} = \mathbf{V}^{H}_{i} \mathbf{Q}_{i} \mathbf{V}_{i}$ and $\hat{\mathbf{g}}_{i} = \mathbf{V}^{H}_{i} \mathbf{g}_{i}$, the optimization problem can be written as
\begin{equation}\label{eq:problem_Qhat}
\begin{array}{ll}
\begin{split}
\mathop{\text{maximize}}_{\substack{\hat{\mathbf{Q}}_i}} 
\end{split}  
& \log\left| \mathbf{I} + \mathbf{\Sigma}_{i} \hat{\mathbf{Q}}_{i} \mathbf{\Sigma}_{i}^{H} \right|\\
\mathrm{subject~to} & \hat{\mathbf{Q}}_{i} \succeq \mathbf{0} \\
& \mathrm{Tr}(\hat{\mathbf{Q}}_i) \leq P_{i} \\
& \hat{\mathbf{g}}_{i}^{H} \hat{\mathbf{Q}}_{i} \hat{\mathbf{g}}_{i} \geq \Gamma_{i}.
\end{array}
\end{equation}
The Hadamard's inequality states that the determinant of a positive definite matrix is less than or equal to the product of its diagonal entries.  Equality in Hadamard's inequality is achieved if and only if the positive definite matrix is diagonal.  Therefore, the optimal $\hat{\mathbf{Q}}_{i}$ of \eqref{eq:problem_Qhat}  has to be diagonal.  With a diagonal matrix $\hat{\mathbf{Q}}_{i}$, the optimization problem \eqref{eq:problem_Qhat} is equivalently
\begin{equation}\label{eq:problem_qhat}
\mathcal{P}_{2}: \begin{array}{ll}
\begin{split}
\mathop{\text{maximize}}_{\substack{\{\hat{q}_{im}\}}}
\end{split}  
& \sum_{m=1}^{M_{t}} \log\left( 1 + |\sigma_{im}|^{2} \hat{q}_{im} \right)\\
\mathrm{subject~to} & \hat{q}_{im} \geq 0 , \,\,\, \forall m \\
& \sum_{m=1}^{M_{t}} \hat{q}_{im} \leq P_{i} \\
& \sum_{m=1}^{M_{t}} |\hat{g}_{im}|^{2} \hat{q}_{im} \geq \Gamma_{i}
\end{array}
\end{equation}
where $\{\hat{q}_{im}\}$ are the diagonal elements of $\hat{\mathbf{Q}}_{i}$, $\{\sigma_{im}\}$ are the diagonal elements of $\mathbf{\Sigma}_{i}$, and $\{\hat{g}_{im}\}$ are the elements of $\hat{\mathbf{g}}_{i}$.
Problem $\mathcal{P}_{2}$ is local to transmitter $i$.  In the following, we remove index $i$ as we understand it is a response of player $i$, i.e., transmitter $i$.  

If $ \max_{m}(|\hat{g}_{m}|^{2}) < \Gamma/P $, Problem $\mathcal{P}_{2}$ has no feasible solution. Otherwise, it can be solved, and the optimal $\{\hat{q}_{m}\}$ are given by a multi-level water-filing solution as
\begin{equation}\label{eq:multilevel-waterfilling}
\hat{q}_{m}^{*} = \max \left\{0,  \frac{1}{\nu_{1} - \nu_{2} |\hat{g}_{m}|^{2}} - \frac{1}{|\sigma_{m}|^{2}} \right\}
\end{equation}
where $\nu_{1}$ and $\nu_{2}$ are the parameters that make $\{\hat{q}_{m}^{*}\}$ satisfy the transmit power constraint and the energy harvesting requirement.  The proof of \eqref{eq:multilevel-waterfilling} is given in Appendix~\ref{ap:Problem2}.

The multi-level water-filling solution has the following properties.  (1) For a particular $m$ that $0 < \nu_{1} - \nu_{2} |\hat{g}_{m}|^{2} < |\sigma_{m}|^{2}$, the water level is $1/(\nu_{1} - \nu_{2}|\hat{g}_{m}|^{2})$. The water level increases when the transmit power limit $P$ increases.  (2) For larger $|\hat{g}_{m}|^{2}$, the $m$th water level $1/(\nu_{1} - \nu_{2}|\hat{g}_{m}|^{2})$ is higher, given that $0 < \nu_{1} - \nu_{2} |\hat{g}_{m}|^{2} < |\sigma_{m}|^{2}$.  
(3) If $\sum_{m=1}^{M_{t}} |\hat{g}_{m}|^{2} \hat{q}_{m} > \Gamma$, then $\nu_{2} = 0$.  This reverts to the conventional optimization problem with the single-level water-filling solution.  With this solution of $\{\hat{q}_{m}\}$, the energy harvesting requirement is already satisfied.

For a relatively large $\Gamma$, as long as $ \Gamma \leq \max_{m}(|\hat{g}_{m}|^{2}) P $, there is a feasible multi-level water-filling solution with positive $\nu_{1}$ and $\nu_{2}$.  
As $\nu_{1} > 0$ and $\nu_{2} > 0$, both the transmit power constraint and the energy harvesting constraint attain equality.  
Suppose $\mathcal{M}$ is the set that contains the spatial directions in which the transmitter transmits, i.e., $\hat{q}_{m} > 0, \forall m \in \mathcal{M}$.  For a feasible solution, $\mathcal{M}$ is not an empty set.  
Using \eqref{eq:multilevel-waterfilling}, we have
\begin{eqnarray}
\sum_{m \in \mathcal{M}} \frac{1}{\nu_{1} - \nu_{2}|\hat{g}_{m}|^{2}} &=& P + \sum_{m \in \mathcal{M}} \frac{1}{|\sigma_{m}|^{2}} = P' \\
\sum_{m \in \mathcal{M}} \frac{|\hat{g}_{m}|^{2}}{\nu_{1} - \nu_{2}|\hat{g}_{m}|^{2}} &=& \Gamma + \sum_{m \in \mathcal{M}} \frac{|\hat{g}_{m}|^{2}}{|\sigma_{m}|^{2}} = \Gamma'
\end{eqnarray}
It follows that
\begin{equation}\label{eq:nu1nu2}
\nu_{1} P' - \nu_{2} \Gamma' = |\mathcal{M}|
\end{equation}
where $|\mathcal{M}|$ is the cardinality of set $\mathcal{M}$.  Replacing $\nu_{1}$, we have
\begin{equation}\label{eq:nu2}
\sum_{m \in \mathcal{M}} \frac{1}{|\mathcal{M}| + \nu_{2} \alpha_{m}} = 1
\end{equation}
where $\alpha_{m} = \Gamma' - |\hat{g}_{m}|^{2} P'$. 
Therefore, $\nu_{2}$ is a positive root of \[\prod_{m}(|\mathcal{M}| + \nu_{2} \alpha_{m}) - \sum_{k=1}^{|\mathcal{M}|}\prod_{m \neq k}(|\mathcal{M}| + \nu_{2} \alpha_{m}). \]

For example, if $|\mathcal{M}| = 2$, $\nu_{2} =  - \sum_{m} (1/\alpha_{m})$.

If $|\mathcal{M}| = 3$, $\nu_{2} =  - \sum_{m} (1/\alpha_{m}) + \left(\sum_{m}(1/\alpha_{m}^{2}) - \sum_{m} \alpha_{m} / \prod_{m}\alpha_{m} \right)^{1/2}$. 

When $|\mathcal{M}| \geq 4$, $\nu_{2}$ can be found using numerical methods.  

The multi-level water-filling solution \eqref{eq:multilevel-waterfilling} of Problem $\mathcal{P}_{2}$ can be calculated as in Algorithm~\ref{alg:multilevel_waterfilling}.
Once the optimal $\{\hat{q}_{m}^{*}\}$ are found, the optimal $\hat{\mathbf{Q}}_{i}^{*}$ is constructed as $\hat{\mathbf{Q}}_{i}^{*} = \textrm{diag}\{\hat{q}_{1}^{*}, \hat{q}_{2}^{*}, \ldots, \hat{q}_{M_{t}}^{*}\}$, and the optimal transmit covariance matrix for user $i$ is $\mathbf{Q}^{*}_{i} = \mathbf{V}_{i} \hat{\mathbf{Q}}_{i}^{*} \mathbf{V}_{i}^{H}$.

\begin{algorithm}[h]
\caption{Multi-level Water-filling Algorithm}\label{alg:multilevel_waterfilling}
\begin{algorithmic}
\item[0.] If $ \max_{m}(|\hat{g}_{m}|^{2}) < \Gamma/P $, there is no feasible solution for $\{\hat{q}_{m}\}$, Stop.
\item[1.] Suppose $\nu_{2} = 0$, find optimal $\{\hat{q}_{m}\}$ as the conventional single-level water-filling solution with transmit power constraint $P$;  
\item[2.] With these calculated $\{\hat{q}_{m}\}$, if energy harvesting requirement $\Gamma$ is satisfied, Exit.  Otherwise, $\nu_{1} > 0$ and $\nu_{2} > 0$. Both the transmit power constraint and the energy harvesting constraint attain equality;
\item[3.] Set a full set $\mathcal{M}$ with all the transmit directions $\mathcal{M} = \{1, 2, \ldots, M_{t}\}$;
\item[4.] Find $\nu_{2}$ using \eqref{eq:nu2}, and find $\nu_{1}$ using  \eqref{eq:nu1nu2};
\item[5.] With $\nu_{1}$ and $\nu_{2}$, calculate $\{\hat{q}_{m}\}$ using \eqref{eq:multilevel-waterfilling};
\item[6.] If $\hat{q}_{m} > 0, \forall m \in \mathcal{M}$, Exit.  Otherwise, if $\hat{q}_{k} = 0, \exists k \in \mathcal{M}$, exclude $k$ from set $\mathcal{M}$.  Go to Step 4.
\end{algorithmic}
\end{algorithm}



\subsection{Game Implementation and Best-Response Dynamics}

How do the $K$ wireless transmitters play the non-cooperative game?  First, transmitter $i$ ($i \in \mathcal{K}$) estimates its user channel $\mathbf{H}_{ii}$ and the vector channel $\mathbf{g}_{i}$ to the RF energy harvester.  (Appendix~\ref{ap:estimate_g} gives a discussion on estimating channel $\mathbf{g}_{i}$.)  Secondly, transmitter $i$ measures the multi-user interference and noise $\mathbf{R}_{i}$.  If transmitter $i$ senses that other transmitters have fixed their transmissions, it reasons that it is its turn to update the transmit covariance matrix.  Thirdly, transmitter $i$ turns off the transmission and acquires the feedback from the RF energy harvester about the partial minimum energy requirement  $\Gamma_{i}$.  Finally, transmitter $i$ solves local Problem $\mathcal{P}_{2}$ with the optimal transmit covariance matrix $\mathbf{Q}_{i}^{*}$.  This is the best response of user $i$ to other users' transmissions regarding maximum information rate.

As discussed before, Game $\Game$ has at least one pure-strategy NE.  If player $i, \forall i \in \mathcal{K}$, has a single best response to each strategy profile $\mathbf{Q}_{-i}$ of the other players, the $K$ strategies at the NE can be put in the collection of best-response equations.  Unfortunately, each player's best response can not be expressed in a single equation.  The NE can only be approached by best-response iterations among the players.  The network iterations can be done with asynchronous player updates.  

Without the energy harvesting requirement, a similar multiuser MIMO transmission game is addressed~\cite{4604735}.  The water-filling operator is interpreted as a projection onto a proper polyhedral set.  Then, a set of sufficient conditions are derived that guarantee the uniqueness of the NE and the global convergence of the distributed asynchronous water-filling algorithms.  The sufficient conditions are based on the user channel $\{\mathbf{H}_{ii}\}_{i \in \mathcal{K}}$ and the interference channels $\{\mathbf{H}_{ij}\}_{i \neq j}$.  In our problem, the uniqueness conditions would also depend on the vector channels $\{\mathbf{g}_{i}\}_{i \in \mathcal{K}}$ to the RF energy harvester.  In general, the NE may not be unique.  Also, the best-response dynamic may cycle in the game and not converge to a NE~\cite{TwentyAlgorithmicGameTheory}.

\section{Wireless Transmissions in a Cooperative Game}
\label{sec:cooperativegame}

\subsection{Cooperative Game}

There are two practical concerns of the non-cooperative game of user transmissions.  First, the proportion of energy contribution imposed on each transmitter may not be optimal.  User $i$ uses the transmission on-and-off process and the energy harvester feedback to estimate the partial minimum energy requirement $\Gamma_{i}$.  It gives little consideration to what is the best distribution of energy contribution among the transmitters.  Furthermore, if other transmitters currently contribute little to the RF energy harvester and the energy harvesting requirement is harsh, user $i$ may just give up transmission by not having a feasible solution of its local optimization problem.  Second, even though there exists a pure NE of the non-cooperative game, it may not be a globally optimal solution for the user information rates.  
The NE point is not necessarily Pareto-efficient and may not be desirable of the network.  Therefore, we do not further pursue the uniqueness condition of the NE or the convergence condition of the non-cooperative game.

For multiple players in a strategic game, the optimal network outcome depends on competition as well as cooperation.  Each player is an autonomous agent, but negotiation with moderate signaling among the players is allowed.  Transmitter $i$ of the game bargain over its share of energy contribution $\Gamma_{i}$ to the RF energy harvester while maximizing its information rate $r_{i}$.  The goal of the cooperative game is to achieve the user information rates that approach the Pareto optimal rates.

For Game $\Game$, the strategies of all the players $\{\mathbf{Q}_{i}\}$ are intertwined to satisfy the energy harvesting requirement.  Therefore, the players need to cooperate with each other for a good distribution of this requirement.  As the transmitters bargain over the assignments of the energy requirement, a bargaining problem can be leveraged to analyze player cooperation~\cite{peters1992axiomatic}.    Game $\Game$ can be regarded as a bargain among the $K$ transmitters over their shares of energy contribution to the RF energy harvester.  To be fair, the unit-reward of the harvested energy should be the same regardless of which transmitter does the RF radiation come from.  In our cooperative game, we set the disagreement point at zero.   Therefore, any rational player tends to cooperate with others~\cite{peters1992axiomatic}.

Several network utility functions can be selected for the cooperative game.  As long as the network utility function is a quasi-concave and strict increasing function of the individual utilities, there is a unique bargaining solution.   For example, the product of the individual utilities can be chosen as the network utility function.  It is called the Nash product and the bargaining solution is the Nash bargaining solution.   In our problem, instead of the Nash bargaining solution, we are interested in finding the Utilitarian solution that maximizes the sum of the individual utilities.  The Utilitarian solution is a Pareto-efficient solution.  It is directly related to the maximum sum rate of the network with some tradeoff in rate fairness among the users.

For Game $\Game$, finding the Utilitarian solution that maximizes the network utility can be formulated as the following optimization problem.
\begin{equation}\label{eq:problem3}
\mathcal{P}_{3}: \begin{array}{ll}
\begin{split}
\mathop{\text{maximize}}_{\substack{\{\mathbf{Q}_i}\}_{i \in \mathcal{K}}} 
\end{split}  
& \sum_{i \in \mathcal{K}} r_{i}(\mathbf{Q}_{i},\mathbf{Q}_{-i})\\
\mathrm{subject~to} & \mathbf{Q}_i \succeq \mathbf{0}, ~~\forall i \in \mathcal{K} \\
& \mathrm{Tr}(\mathbf{Q}_i) \leq P_i, ~~\forall i \in \mathcal{K}\\
& \sum_{i \in \mathcal{K}} \mathbf{g}_{i}^{H} \mathbf{Q}_{i} \mathbf{g}_{i} \geq \Gamma.
\end{array}
\end{equation}
In the cooperative game, the distribution of the energy harvesting requirement is not pre-determined.  The players adjust their transmit strategies to maximize the network utility as well as reach an agreement of allocating the energy contributions.   In general, maximizing the sum information rate is a non-convex problem and is hard to solve.  

\subsection{Problem Decomposition}

As the transmitters play the cooperative game, the strategy of transmitter $i$ ($i \in \mathcal{K}$) is the transmit covariance matrix $\mathbf{Q}_{i}$.  With Problem $\mathcal{P}_{3}$ on the maximization of the sum information rate, the players can bargain over the last constraint, i.e., the energy harvesting requirement.  A distributed algorithm is desirable for implementing the bargaining process among the players.   The network utility is the sum information rate which is not concave in $\{\mathbf{Q}_{i}\}_{i \in \mathcal{K}}$.  Before we can decompose the optimization problem into several subproblems for the cooperative game, we need to modify the network utility so that it is concave in $\{\mathbf{Q}_{i}\}_{i \in \mathcal{K}}$.

To facilitate the distributed bargaining among the players, we reconstruct the individual utility function from an approximation of the maximum information rate. 
The maximum information rate $r_{i}$ can be approximated with a first-order Taylor expansion around $\mathbf{R}_i = \tilde{\mathbf{R}}_i$ as
\begin{eqnarray} \label{eq:rate_approximate}
r_{i}(\mathbf{Q}_{i}, \mathbf{Q}_{-i}) 
& \approx & \log \left| \mathbf{I} + \tilde{\mathbf{R}}_i^{-1}\mathbf{H}_{ii}\mathbf{Q}_i\mathbf{H}_{ii}^H \right| \nonumber \\
&& - \mathrm{Tr}\left((\tilde{\mathbf{R}}_i^{-1} - (\tilde{\mathbf{R}}_i+\mathbf{H}_{ii}\mathbf{Q}_i\mathbf{H}_{ii}^H)^{-1}) \right. \nonumber \\
&& \left. (\mathbf{R}_i-\tilde{\mathbf{R}}_i)\right) \nonumber \\
&\approx & \log \left| \mathbf{I} + \tilde{\mathbf{R}}_i^{-1}\mathbf{H}_{ii}\mathbf{Q}_i\mathbf{H}_{ii}^H \right| - \mathrm{Tr}(\mathbf{A}_{i} \mathbf{R}_{i}) + c \nonumber \\ \label{eq:approximation1} 
\end{eqnarray}
where $\mathbf{A}_i = \tilde{\mathbf{R}}_i^{-1} - (\tilde{\mathbf{R}}_i+\mathbf{H}_{ii}\tilde{\mathbf{Q}}_i\mathbf{H}_{ii}^H)^{-1}$ and $c=\mathrm{Tr} (\mathbf{A}_{i}\tilde{\mathbf{R}}_{i} )$.  The derivation of the approximation in \eqref{eq:rate_approximate} is shown in Appendix~\ref{ap:TaylorAppro}.   Notice that we replace $\mathbf{Q}_{i}$ with $\tilde{\mathbf{Q}}_{i}$ in $\mathbf{A}_{i}$ as another approximation.

To achieve a good expansion neighborhood that reduces the approximate error of the Taylor polynomial, $\tilde{\mathbf{R}}_{i}$ can be calculated as
$
\tilde{\mathbf{R}}_{i} = \sum_{j \neq i} \mathbf{H}_{ij}\tilde{\mathbf{Q}}_{j}\mathbf{H}_{ij}^H + \mathbf{I}
$, 
and $\tilde{\mathbf{Q}}_{i}$ is the solution of the following convex optimization problem
\begin{equation}\label{eq:Q_tilde}
\begin{array}{ll}
\begin{split} 
\mathop{\text{maximize}}_{\substack{\mathbf{Q}_i}} 
\end{split}  
& \log \left| \mathbf{I} + \mathbf{H}_{ii}\mathbf{Q}_i\mathbf{H}_{ii}^H \right|\\
\mathrm{subject~to} & \mathbf{Q}_i \succeq \mathbf{0} \\
& \mathrm{Tr}(\mathbf{Q}_i) \leq P_{i} \\
& \mathbf{g}_{i}^{H} \mathbf{Q}_i \mathbf{g}_{i} \geq \frac{\Gamma}{K}.
\end{array}
\end{equation}
Here, the energy harvesting requirement is evenly undertaken by the $K$ transmitters.  This is assumed only to estimate a point around which the maximum information rate is Taylor-expanded. Later, the optimal transmissions are determined with a deliberate distribution of the energy harvesting requirement among the $K$ users.


Let us define the utility function of transmitter $i$ ($i \in \mathcal{K}$) as
\begin{equation} \label{eq:utility_game3}
u_{i}(\mathbf{Q}_i) =  \log \left| \mathbf{I} + \tilde{\mathbf{R}}_i^{-1}\mathbf{H}_{ii}\mathbf{Q}_i\mathbf{H}_{ii}^H \right|- \sum_{j \neq i}\mathrm{Tr}(\mathbf{A}_{j} \mathbf{H}_{ji}\mathbf{Q}_{i}\mathbf{H}_{ji}^H).
\end{equation}
Comparing \eqref{eq:utility_game3} with \eqref{eq:approximation1}, we notice that the individual utility function comes from the approximation of the maximum information rate.  Moreover, the second terms in \eqref{eq:approximation1}, i.e., $-\mathrm{Tr}(\mathbf{A}_{i} \sum_{i \neq j} \mathbf{H}_{ij} \mathbf{Q}_{j} \mathbf{H}_{ij}^{H})$, are interchanged among the players and become the second terms in \eqref{eq:utility_game3}. The second part of the utility function  $u_{i}(\mathbf{Q}_{i})$ can be regarded as the penalty for player $i$ due to the interference it causes to other users.  
 The utility function $u_{i}(\mathbf{Q}_{i})$ only depends on transmit covariance matrix $\mathbf{Q}_{i}$, and it is concave in $\mathbf{Q}_{i}$.  Therefore, $\sum_{i \in \mathcal{K}} u_{i}(\mathbf{Q}_{i})$ is concave in $\{\mathbf{Q}_{i}\}_{i \in \mathcal{K}}$.
This makes it possible for a distributed implementation of the cooperative game.  The cost of this method is that player $i$ needs to know the cross-channel matrices $\{\mathbf{H}_{ji}\}_{j \in \mathcal{K} \setminus \{i\}}$.  
The interference channel $\mathbf{H}_{ji}$ can be estimated and fed back to transmitter $i$ through user cooperation.  

Problem $\mathcal{P}_{3}$ now becomes 
\begin{equation}\label{eq:problem4}
\mathcal{P}_{4}: \begin{array}{ll}
\begin{split}
\mathop{\text{maximize}}_{\substack{\{\mathbf{Q}_i}\}_{i \in \mathcal{K}}} 
\end{split}  
& \sum_{i \in \mathcal{K}} u_{i}(\mathbf{Q}_{i})\\
\mathrm{subject~to} & \mathbf{Q}_i \succeq \mathbf{0}, ~~\forall i \in \mathcal{K} \\
& \mathrm{Tr}(\mathbf{Q}_i) \leq P_i, ~~\forall i \in \mathcal{K}\\
& \sum_{i \in \mathcal{K}} \mathbf{g}_{i}^{H} \mathbf{Q}_{i} \mathbf{g}_{i} \geq \Gamma.
\end{array}
\end{equation}
Problem $\mathcal{P}_{4}$ is a convex optimization problem with a coupling constraint on the strategies $\{\mathbf{Q}_{i}\}_{i \in \mathcal{K}}$.  We use the dual decomposition method to decompose the Lagrangian dual problem~\cite{palomar_tutorial_2006,NonlinearProgramming}.  
We decompose the problem into subproblems for the players and develop a distributed algorithm to find the bargaining solution of the cooperative game.

The Lagrangian of Problem $\mathcal{P}_{4}$ is given by
\begin{eqnarray}
L(\{\mathbf{Q}_{i}\}, \lambda) 
&=&\sum_{\substack{i \in \mathcal{K} }} \log \left|\mathbf{I} + \tilde{\mathbf{R}}_i^{-1}\mathbf{H}_{ii}\mathbf{Q}_i\mathbf{H}_{ii}^H \right| \nonumber \\
&& - \sum_{i \in \mathcal{K}} \sum_{j \neq i}\mathrm{Tr}\left(\mathbf{A}_{j} \mathbf{H}_{ji}\mathbf{Q}_{i}\mathbf{H}_{ji}^H \right)  \nonumber \\
&&  - \lambda \left(\Gamma -  \sum_{i \in \mathcal{K}} \mathbf{g}_{i}^{H} \mathbf{Q}_{i} \mathbf{g}_{i} \right) \label{eq:lagrangian}
\label{eq:Game1 lagrangian}
\end{eqnarray}
with $\mathbf{Q}_i \succeq \mathbf{0}$ and $\mathrm{Tr}(\mathbf{Q}_i) \leq P_i$, $\forall i \in \mathcal{K}$.  $\lambda$ is the dual variable associated with the energy harvesting requirement.
With the dual decomposition method, the network utility maximization problem can be divided into two levels of optimization problems.  At the lower level, each player solves the local convex optimization problem as
\begin{equation}
\begin{array}{ll}
\begin{split}
\mathop{\mathrm{maximize}}_{\substack{\mathbf{Q}_i}} 
\end{split} 
&\log \left| \mathbf{I} + \tilde{\mathbf{R}}_i^{-1}\mathbf{H}_{ii}\mathbf{Q}_i\mathbf{H}_{ii}^H \right| \\
& - \sum_{j \neq i}\mathrm{Tr}\left(\mathbf{A}_{j}\mathbf{H}_{ji}\mathbf{Q}_{i}\mathbf{H}_{ji}^H \right) + \lambda \mathbf{g}_{i}^{H} \mathbf{Q}_{i}\mathbf{g}_{i}\\
\mathrm{subject~to} &\mathbf{Q}_i \succeq \mathbf{0} \\
& \mathrm{Tr}(\mathbf{Q}_i) \leq P_i.
\end{array}
\label{eq:local_problem}
\end{equation}
The energy contribution of transmitter $i$ to the RF energy harvester is denoted as $\beta_{i} = \mathbf{g}_{i}^{H} \mathbf{Q}_{i} \mathbf{g}_{i}$.  $\lambda$ can be interpreted as the unit-reward for energy contribution.  

At the higher level, the master problem updates the dual variable $\lambda$ by solving the dual problem
\begin{equation}
\begin{array}{ll}
\begin{split}
\mathop{\text{minimize}}_{\substack{\lambda}} 
\end{split}  
& g(\lambda) = \sum_{i \in \mathcal{K}} g_{i}(\lambda) - \lambda \Gamma\\
\mathrm{subject~to} & \lambda \geq 0
\end{array}
\end{equation}
where $g_{i}(\lambda)$ is the dual function obtained as the maximum value of the Lagrangian solved in \eqref{eq:local_problem} for a given $\lambda$.  
The decomposition method solves the dual problem instead of the primal problem. Since the original problem is convex and there exist strictly feasible solutions, strong duality holds.
At the higher level, the unit-reward $\lambda$ can be determined by the sub-gradient method~\cite{NonlinearProgramming}. In Game $\Game$, we use the bisection method to obtain $\lambda$.

\subsection{Distributed Bargaining}
\label{sec:distributedbargaining}

With the dual decomposition method at our disposal, we can orchestrate the transmitters to play the cooperative game with distributed bargaining.  During the initialization phase, first, transmitter $i$ ($i \in \mathcal{K}$) estimates its user channel $\mathbf{H}_{ii}$ and the vector channel $\mathbf{g}_{i}$ to the RF energy harvester.  Transmitter $i$ also estimates the cross-channel matrices $\{\mathbf{H}_{ij}\}$.  Secondly, transmitter $i$ finds $\tilde{\mathbf{Q}}_{i}$ as the solution of \eqref{eq:Q_tilde}.  It then calculates $\tilde{\mathbf{R}}_{i}$ and $\mathbf{A}_{i}$.   Alternatively, $\tilde{\mathbf{R}}_{i}$ can be measured by letting the transmitters transmit using $\{\tilde{\mathbf{Q}}_{i}\}$.  Thirdly, the transmitters broadcast $\{\mathbf{A}_{i}\}$ and cross-channels $\{\mathbf{H}_{ij}\}$ to other transmitters in the network.

After the initialization phase, the transmitters solve the lower-level subproblems distributively.  The transmitters set initial values of $\lambda^{(0)}_{\mathrm{min}}$ and $\lambda^{(0)}_{\mathrm{max}}$ and calculate the unit-reward of energy contribution as $\lambda^{(0)} = (\lambda^{(0)}_{\mathrm{min}} + \lambda^{(0)}_{\mathrm{max}})/2$.  Given the unit-reward $\lambda^{(0)}$, each transmitter determines its transmit covariance matrix $\mathbf{Q}_{i}$ by solving the local convex optimization problem \eqref{eq:local_problem}.  The subproblems are coordinated by the high-level master problem through signaling.   The energy contribution $\beta_{i} = \mathbf{g}_{i}^{H} \mathbf{Q}_{i} \mathbf{g}_{i}$ is shared among the players.  After receiving $\{\beta_{i}\}$ from all of the players, each transmitter updates the unit-reward of energy contribution with the bisection rule as
\begin{equation}
\left\{ \begin{array}{cc}
  \lambda^{(z+1)}_{\mathrm{min}} = \lambda^{(z)} & \text{if~~} \sum_{i \in \mathcal{K}} \beta_{i} < \Gamma \\
  \lambda^{(z+1)}_{\mathrm{max}} = \lambda^{(z)} & \text{if~~} \sum_{i \in \mathcal{K}} \beta_{i} \geq \Gamma 
\end{array}   \right. 
\end{equation}
\begin{equation}
\lambda^{(z+1)} = \frac{1}{2}\left(\lambda^{(z+1)}_{\mathrm{min}} +  \lambda^{(z+1)}_{\mathrm{max}} \right)
\end{equation}
where $z = 0, 1, 2, \ldots$ is the iteration index.  For example, if $\sum_{i \in \mathcal{K}} \beta_{i} < \Gamma$, which means that the energy harvesting requirement is not met, the unit-reward $\lambda$ will increase.   This gives an incentive for the transmitters to adjust their transmissions for more contribution to the RF energy harvester.  
The players have the same value $\lambda_{\mathrm{min}}$, the same value $\lambda_{\mathrm{max}}$, and the same bisection rule.  It guarantees that each transmitter generates the same updated $\lambda$.   The players may compare their updated values of $\lambda$ and reach a consensus if there is any discrepancy due to signaling errors.   Alternatively, one representative player can update $\lambda$ and broadcast the new value to others.  In this way, the transmitters effectively bargain over the unit-reward $\lambda$, and it is fair to every player.  With a new $\lambda$, each transmitter determines its transmit covariance matrix again. Solving the local optimization problem does not need new information except for the updated unit-reward $\lambda$.  Each iteration of the algorithm can be regarded as a bargaining round.  After several rounds of bargaining, the algorithm stops when the unit-reward $\lambda$ converges.  That is, a deal is made on the unit-reward and how to distribute the energy contributions among the transmitters.

The distributed bargaining process of the cooperative game is summarized in Algorithm \ref{algorithm}.  Except for the steps in the initialization phase, only scalar-valued energy contributions $\{\beta_{i}\}$ are to be exchanged among the $K$ users.  Therefore, the network signaling can be kept moderate. 

Finally, we discuss how to overcome the approximation error in \eqref{eq:rate_approximate}.
If $\tilde{\mathbf{Q}}_{i}$ initially calculated from \eqref{eq:Q_tilde} is very different from the optimal $\mathbf{Q}_{i}$, the error in the approximation can be relatively large.  This may lead to a sum rate that is much less than the optimal sum rate.  In this case, we can update $\tilde{\mathbf{Q}}_{i}$ with the bargaining solution $\mathbf{Q}_{i}^{*}$, i.e., $\tilde{\mathbf{Q}}_{i} = \mathbf{Q}_{i}^{*}$.   With updated $\{\tilde{\mathbf{Q}}_{i}\}$, $\{\tilde{\mathbf{R}}_{i}\}$ and $\{\mathbf{A}_{i}\}$, we repeat the entire distributed bargain process.  In a few iterations of updating $\tilde{\mathbf{Q}}_{i}$ and repeating the bargaining process, the algorithm provides user achievable information rates that approach the Pareto front.  

\begin{algorithm}[h]
\caption{Distributed Bargaining Process of Cooperative Game}\label{algorithm}
\begin{algorithmic}
\item[0.] Initialization Phase:  Each player estimates $\mathbf{H}_{ii}$, $\mathbf{g}_{i}$, and $\{\mathbf{H}_{ij}\}$.  Each player finds $\tilde{\mathbf{Q}}_{i}$ as the solution of \eqref{eq:Q_tilde} and calculates $\tilde{\mathbf{R}}_{i}$ and $\mathbf{A}_{i}$.   The transmitters send $\{\mathbf{A}_{i}\}$ and $\{\mathbf{H}_{ij}\}$ to other players;
\item[1.] The players set $\lambda^{(0)}_{\mathrm{min}}$ and $\lambda^{(0)}_{\mathrm{max}}$.  $z=0$;
\item[2.] Each player calculates $\lambda^{(z)} = (\lambda^{(z)}_{\mathrm{min}} + \lambda^{(z)}_{\mathrm{max}})/2$;
\item[3.] With the unit-reward of energy contribution $\lambda$, each player determines the optimal transmit covariance matrix $\mathbf{Q}_{i}$ by solving the local problem \eqref{eq:local_problem};
\item[4.] Each player calculates its energy contribution $\beta_{i}$ and broadcasts this value to other players;
\item[5.] According to the bisection rule, each player updates the unit-reward $\lambda$ with all the values of $\{\beta_{i}\}$.  That is, if $\sum_{i \in \mathcal{K}} \beta_{i} < \Gamma$, $\lambda^{(z+1)}_{\mathrm{min}} = \lambda^{(z)}$, otherwise $\lambda^{(z+1)}_{\mathrm{max}} = \lambda^{(z)}$.  $z = z +1$;
\item[6.] Go to Step 2 until $\lambda$ converges, that is, $\lambda_{\mathrm{max}} - \lambda_{\mathrm{min}}$ is smaller than a threshold.
\end{algorithmic}
\end{algorithm}

\subsection{Multiple Energy Harvesters}
\label{sec:multipleharvesters}


When there are multiple RF energy harvesters in the vicinity of the wireless communication transmitters, the transmission algorithms need to be modified to satisfy the energy harvesting requirement of each harvester.  Suppose that the wireless communication systems are aware of the multiple RF energy harvesters.  The transmitters deliberately concentrate the RF radiations to charge this network of energy harvesters.  Any RF energy harvester can dynamically join and leave the energy harvesting network.  An energy harvester requests and joins the network when its battery is depleted and needs a recharge.   An energy harvester leaves the network once it has enough collected energy.

Suppose that $\mathcal{L}$ is the finite set of the $L$ RF energy harvesters that are actively considered by the wireless transmitters.  If a device's battery is critically low, i.e., below a lower threshold, the device registers onto the energy harvesting network.  The wireless communication systems consider its energy harvesting requirement.  If a device's battery is sufficiently charged, i.e., beyond an upper threshold, the device reports and leaves the energy harvesting network.  The wireless communication systems remove this energy harvester from their transmission consideration.  Nevertheless, this device harvests RF energy anyways without the consciousness of the wireless communication systems.  
Upon changes of Set $\mathcal{L}$, i.e., any device joining or leaving the RF energy harvesting network, the wireless communication systems redesign their transmit covariance matrices to satisfy the energy harvesting requirements. 

With multiple RF energy harvesters, the last constraint of Problem $\mathcal{P}_{4}$ is modified as
\begin{equation}
\sum_{i \in \mathcal{K}} \mathbf{g}_{li}^{H} \mathbf{Q}_{i} \mathbf{g}_{li} \geq \Gamma_{l}, ~~\forall l \in \mathcal{L}.
\end{equation}
The energy harvesting requirements are reflected in a set of constraints, with $\Gamma_{l}$ being the energy requirement of the $l$th harvester.  The $i$th transmitter knows the channel vector $\mathbf{g}_{li}$ to the $l$th energy harvester.  The transmit covariance matrix $\mathbf{Q}_{i}$ needs to satisfy $\mathbf{g}_{li}^{H} \mathbf{Q}_{i} \mathbf{g}_{li} \geq \Gamma_{l} - \sum_{k \in \mathcal{K} \setminus \{i\}} \mathbf{g}_{ki}^{H} \mathbf{Q}_{k} \mathbf{g}_{ki} = \Gamma_{li}$, where $\Gamma_{li}$ is the required energy contribution from the $i$th transmitter to the $l$th RF energy harvester.  


Because of the coupling constraints, we use a similar dual decomposition to decompose the convex optimization problem into two levels of subproblems.  At the lower level, each transmitter solves a local convex optimization problem as
\begin{equation}
\begin{array}{ll}
\begin{split}
\mathop{\mathrm{maximize}}_{\substack{\mathbf{Q}_i}} 
\end{split} 
&\log \left| \mathbf{I} + \tilde{\mathbf{R}}_i^{-1}\mathbf{H}_{ii}\mathbf{Q}_i\mathbf{H}_{ii}^H \right|\\ 
& - \sum_{j \neq i}\mathrm{Tr}\left(\mathbf{A}_{j}\mathbf{H}_{ji}\mathbf{Q}_{i}\mathbf{H}_{ji}^H \right) + \sum_{l} \lambda_{l} \mathbf{g}_{li}^{H} \mathbf{Q}_{i}\mathbf{g}_{li}\\
\mathrm{subject~to} &\mathbf{Q}_i \succeq \mathbf{0} \\
& \mathrm{Tr}(\mathbf{Q}_i) \leq P_i
\end{array}
\label{eq:local_problem2}
\end{equation}
where $\lambda_{l}$ is the dual variable associated with the energy harvesting requirement of the $l$th RF energy harvester.
At the higher level, the master problem updates the dual variables $\{\lambda_{l}\}$ by solving the dual problem
\begin{equation}
\begin{array}{ll}
\begin{split}
\mathop{\text{minimize}}_{\substack{\{\lambda_{l}}\}} 
\end{split}  
& g(\{\lambda_{l}\}) = \sum_{i \in \mathcal{K}} g_{i}(\{\lambda_{l}\}) - \sum_{l \in \mathcal{L}} \lambda_{l} \Gamma_{l}\\
\mathrm{subject~to} & \lambda_{l} \geq 0, \,\,\, \forall l \in \mathcal{L}
\end{array}
\end{equation}
where $g_{i}(\{\lambda_{l}\})$ is the dual function obtained as the maximum value of the Lagrangian solved in \eqref{eq:local_problem2} for given $\{\lambda_{l}\}$.  The master dual problem of $g(\{\lambda_{l}\})$ can be solved with a sub-gradient method (see Appendix~\ref{ap:subgradient}).  The decomposition method solves the dual problem, and strong duality holds for the primal problem and the dual problem.


\begin{figure}[!t]
\centering
\includegraphics[width=0.48\textwidth]{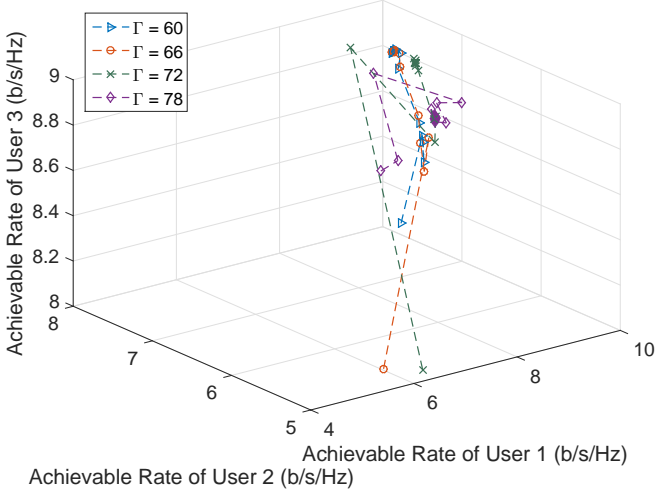}
\caption{Achievable information rates of $K = 3$ wireless users over non-cooperative game iterations.  Channel Set I.}
\label{fig:noncoop_rate3D_Channel1}
\end{figure}

The dual variable $\lambda_{l}$ can be interpreted as the unit-reward for energy contribution to the $l$th energy harvester.  In the master dual problem, the multiple players bargain over the unit-rewards for energy contribution to the energy harvesters.  The unit-rewards $\{\lambda_{l}\}$ can be different for different energy harvesters according to the channels $\{\mathbf{g}_{li}\}$ and the energy requirements $\{\Gamma_{l}\}$.  This creates an incentive for the transmitters to concentrate the RF energy to where it is needed the most.  

Note that, as for future research consideration, the channels $\{\mathbf{g}_{li}\}$ from the wireless communication transmitters to the RF energy harvesters may be fixed but unknown to the transmitters.  To establish a transmission game with RF energy harvesting, the energy harvesters that are attached to IoT devices can feed back their battery charging statuses through the Internet.  Suppose that the feedback delay is negligible.   Energy harvester $l$ can measure its RF energy harvesting rate (battery charging rate) $\psi_{l} = \sum_{i \in \mathcal{K}} \mathbf{g}_{li}^{H} \mathbf{Q}_{i}\mathbf{g}_{li}$ and feed back $\psi_{l}$ to any of the transmitters.  Also, the IoT device with energy harvester $l$ may simply feed back an indication whether $\psi_{l} \geq \Gamma_{l}$ is satisfied.


\section{Numerical Results}
\label{sec:numericalresults}

\begin{figure}[!t]
\centering
\includegraphics[width=0.48\textwidth]{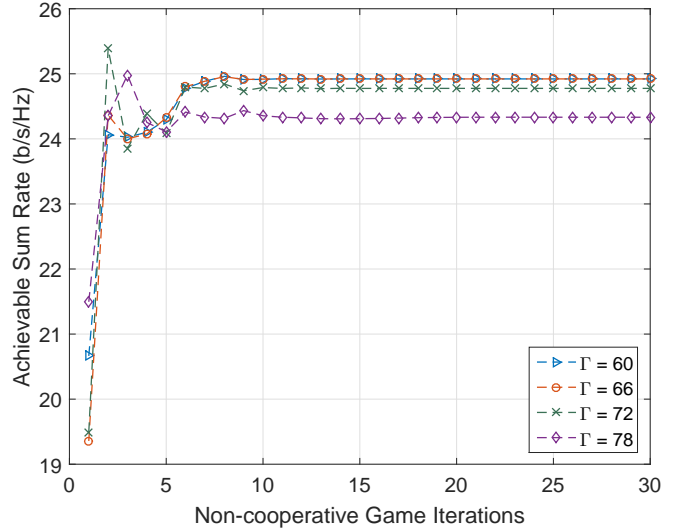}
\caption{Achievable sum rate of $K = 3$ wireless users in a non-cooperative game.  Channel Set I.}
\label{fig:noncoop_sumrate_Channel1}
\end{figure}


We simulate a network of $K =3$ wireless users, i.e., three pairs of wireless transmitter and receiver.  The transmissions of the wireless users interfere with each other.   Each user adjusts its transmit covariance matrix $\mathbf{Q}_{i}$ ($i \in \mathcal{K}$) to maximize the information rate.  In the vicinity of the transmitters, there is an RF energy harvester.  The harvester requires a certain amount of received RF energy collectively from the three wireless transmitters.  Each transmitter has $M_{t} = 8$ transmit antennas and each receiver has $M_{r} = 8$ receive antennas, hence $8 \times 8$ MIMO channels.  The RF energy harvester has one receive antenna.   The user channel matrices $\mathbf{H}_{ii}$  ($i \in \mathcal{K}$), the interference channel matrices $\mathbf{H}_{ij}$ ($ i \neq j$), and the channel vectors toward the energy harvester $\mathbf{g}_{i}$ ($ i \in \mathcal{K}$) are arbitrary but do not vary over a long period of user transmission.  We randomly select channel set $\{\mathbf{H}_{ii}, \mathbf{H}_{ij}, \mathbf{g}_{i}\}$ to simulate the wireless channels for the transmission game.  Each element of the channels is independent circularly symmetric complex Gaussian with zero mean and unit variance.  The transmit power constraints are set as $P_{i} = 8$ ($ \forall i \in \mathcal{K}$) and the energy harvesting requirement $\Gamma$ is around the range $[40, 90]$ so that the optimization problem is feasible and the energy harvesting constraint is effective.

\begin{figure}[!t]
\centering
\includegraphics[width=0.48\textwidth]{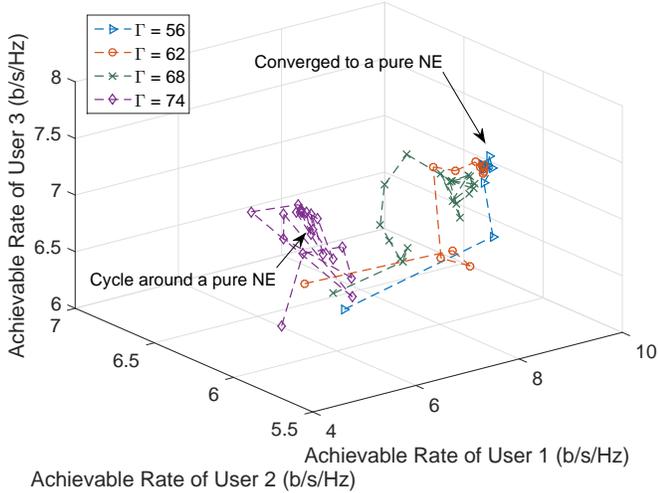}
\caption{Achievable information rates of $K = 3$ wireless users over non-cooperative game iterations.  Channel Set II.}
\label{fig:noncoop_rate3D_Channel2}
\end{figure}
\begin{figure}[!t]
\centering
\includegraphics[width=0.48\textwidth]{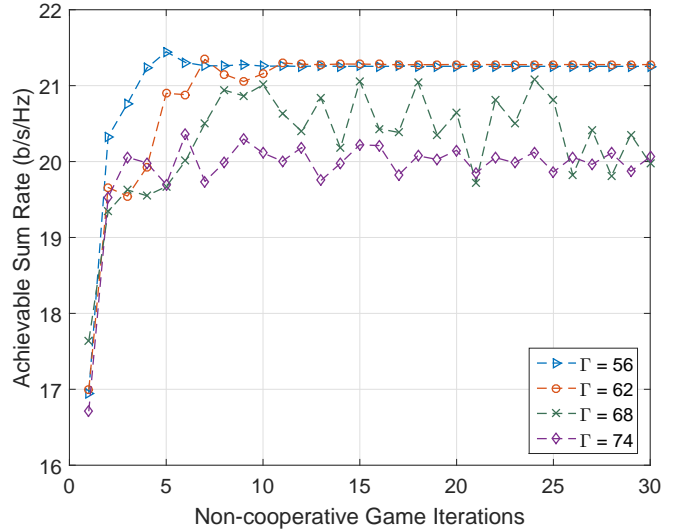}
\caption{Achievable sum rate of $K = 3$ wireless users in a non-cooperative game.  Channel Set II.}
\label{fig:noncoop_sumrate_Channel2}
\end{figure}

When a non-cooperative game is played among the three wireless transmitters, the transmitters start with independent initial guesses on transmit covariance matrices and update them sequentially.  At its turn, transmitter $i$ measures the interference and noise $\mathbf{R}_{i}$, acquires the partial minimum energy requirement $\Gamma_{i}$, and solves the local optimization problem $\mathcal{P}_{2}$ with the multi-level water-filing solution.   This solution is the best response of user $i$ to others' transmission strategies that maximizes the information rate unilaterally.  With an arbitrary Channel Set I, Fig.~\ref{fig:noncoop_rate3D_Channel1} shows the achievable information rates of the three users over game iterations.  Each iteration includes a round of sequential responses of the three transmitters.   As $\Gamma = 70$, the game process converges to a pure NE.   As $\Gamma = 90$, the game process cycles around a pure NE.    Fig.~\ref{fig:noncoop_sumrate_Channel1} shows the achievable sum rate of the three users over game iterations, where the energy harvesting requirement $\Gamma$ ranges from 50 to 90.    With a more demanding $\Gamma$, the achievable sum rate suffers and the best-response dynamic may cycle around a pure NE.  With another arbitrary Channel Set II, Fig.~\ref{fig:noncoop_rate3D_Channel2} and Fig.~\ref{fig:noncoop_sumrate_Channel2} show the achievable information rates and the achievable sum rate, respectively, of the three users over game iterations.   The energy harvesting requirement $\Gamma$ ranges from 40 to 80.  Similar effects on the achievable rates are revealed.

\begin{figure}[!t]
\centering
\includegraphics[width=0.48\textwidth]{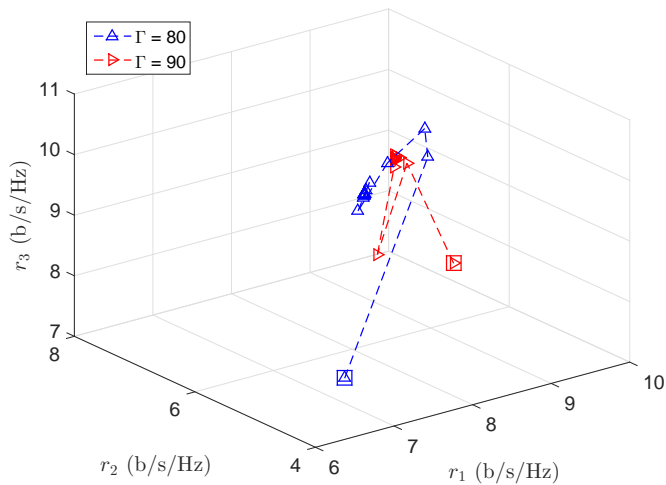}
\caption{Achievable information rates of $K=3$ wireless users in a cooperative game over bargaining iterations.  The starting points are indicated with $\square$.  Channel Set I.}
\label{fig:rate3D_Channel1_bargain}
\end{figure}

\begin{figure}[!t]
\centering
\includegraphics[width=0.48\textwidth]{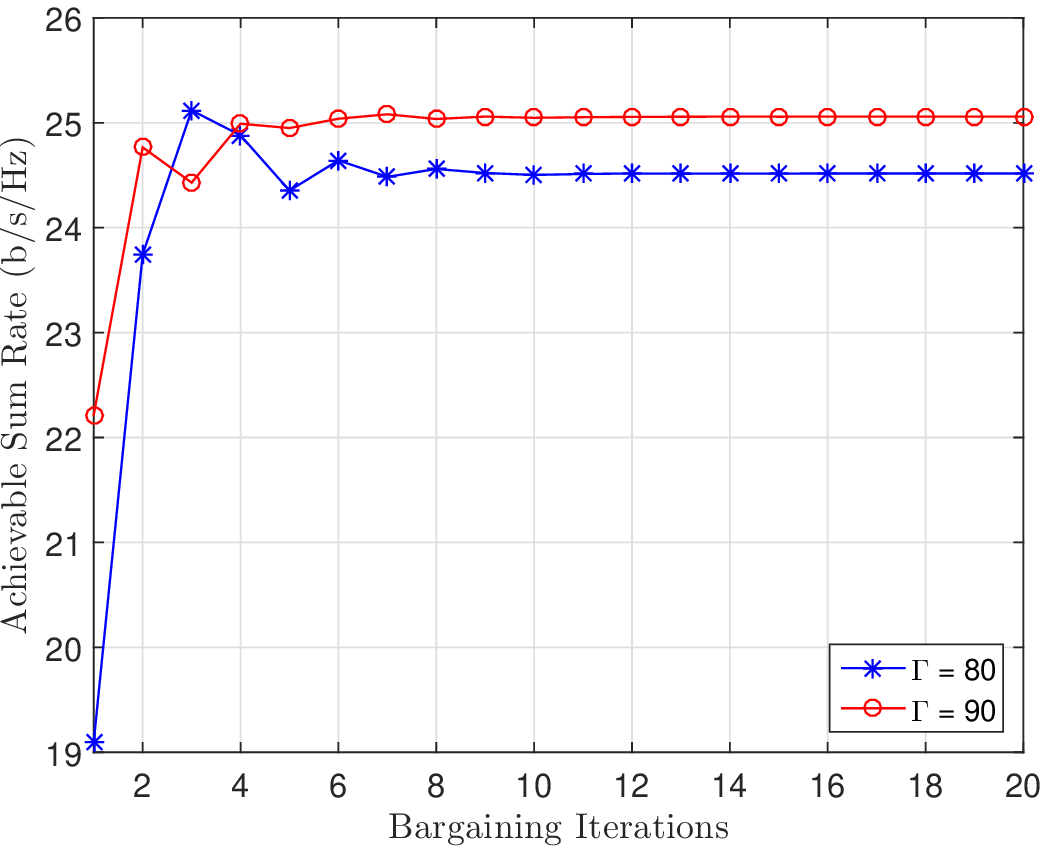}
\caption{Achievable sum rate of $K=3$ wireless users in a cooperative game over bargaining iterations.  Channel Set I.}
\label{fig:sumrate_Channel1_bargain}
\end{figure}

\begin{figure}[!t]
\centering
\includegraphics[width=0.48\textwidth]{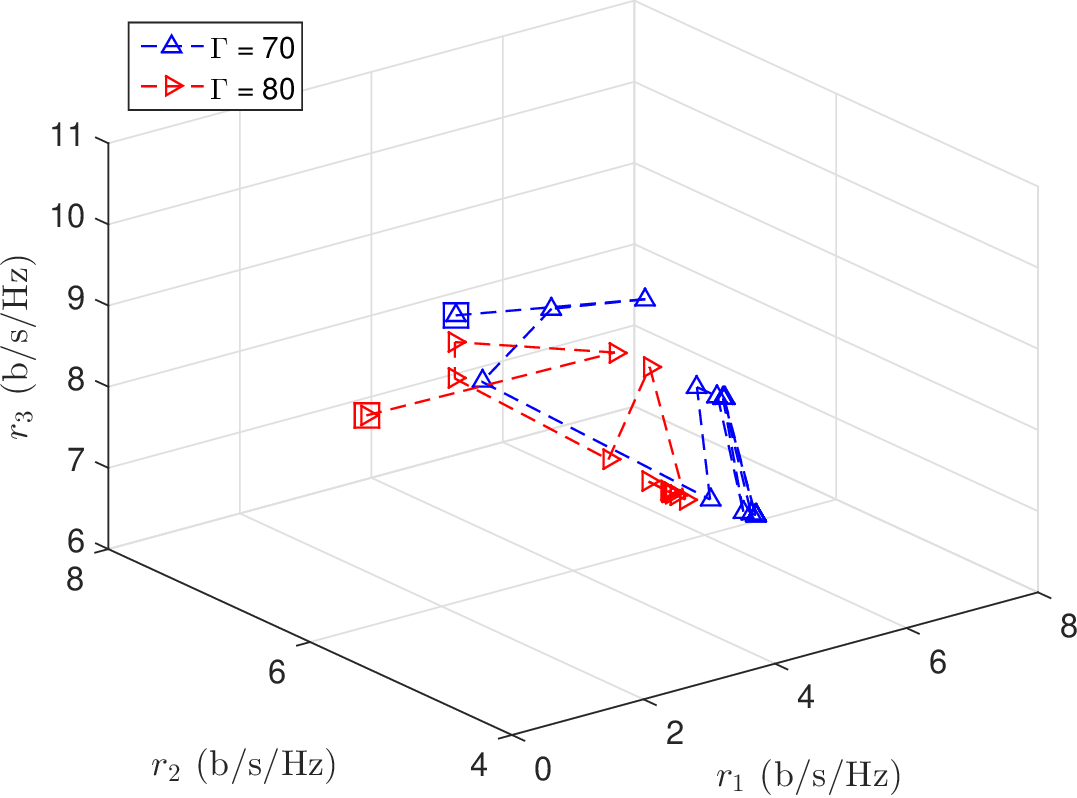}
\caption{Achievable information rates of $K=3$ wireless users in a cooperative game over bargaining iterations.  The starting points are indicated with $\square$.  Channel Set II.}
\label{fig:rate3D_Channel2_bargain}
\end{figure}

\begin{figure}[!t]
\centering
\includegraphics[width=0.48\textwidth]{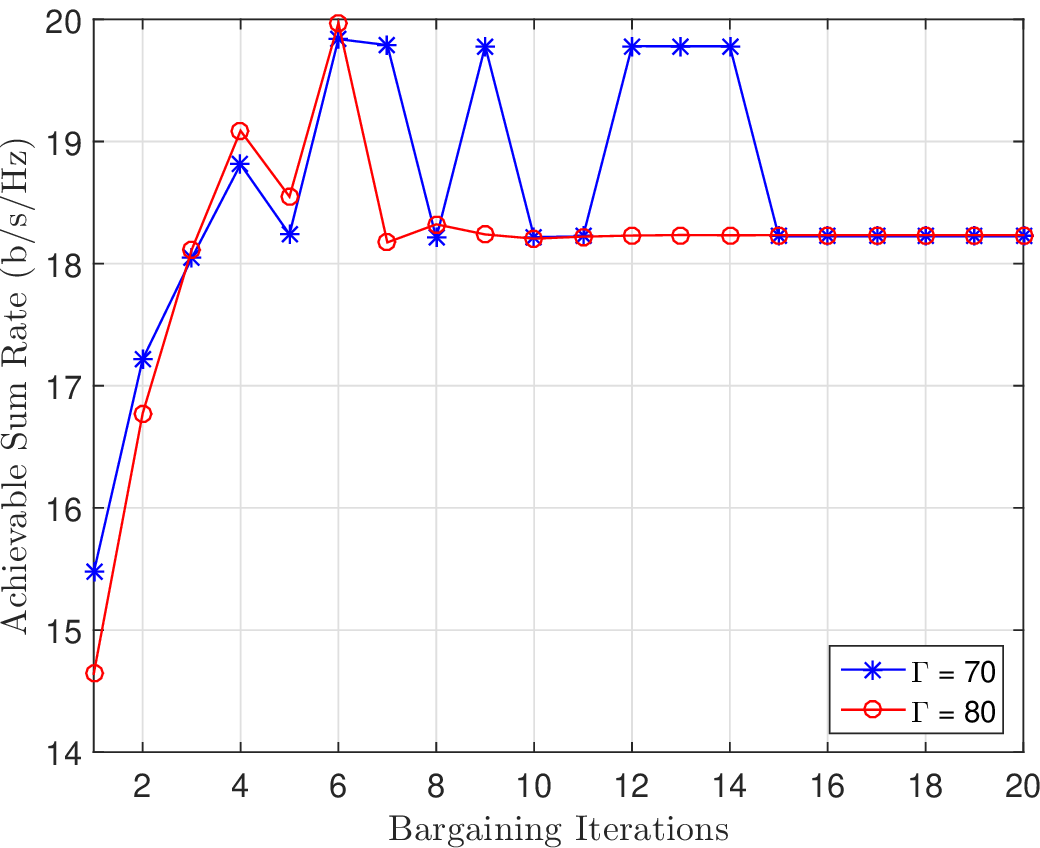}
\caption{Achievable sum rate of $K=3$ wireless users in a cooperative game over bargaining iterations.  Channel Set II.}
\label{fig:sumrate_Channel2_bargain}
\end{figure}

With Channel Set I and Set II the same as above, a cooperative game is played among the three wireless transmitters to maximize the sum utility.  The individual utility is derived from an approximation of the achievable information rate.  A distributed bargaining process is implemented in the network to solve optimization problem $\mathcal{P}_{4}$.  The three users bargain sequentially over the energy contributions and finally agree on the unit-reward $\lambda$.  With Channel Set I, Fig.~\ref{fig:rate3D_Channel1_bargain} and Fig.~\ref{fig:sumrate_Channel1_bargain} show the achievable information rates and the achievable sum rate, respectively, of the three users over bargaining iterations.   The initial $\tilde{\mathbf{Q}}_{i}$ is found as the solution of \eqref{eq:Q_tilde}.   With Channel Set II, Fig.~\ref{fig:rate3D_Channel2_bargain} and Fig.~\ref{fig:sumrate_Channel2_bargain} show the achievable information rates and the achievable sum rate, respectively, of the three users over bargaining iterations.   If $\tilde{\mathbf{Q}}_{i}$ is close to the optimal $\mathbf{Q}_{i}$, the approximation error of the utility function is negligible.  The bargaining process has better performance than the non-cooperative game.   This is the case with Channel Set I, as we compare the  corresponding sum rates in Fig.~\ref{fig:sumrate_Channel1_bargain} and Fig.~\ref{fig:noncoop_sumrate_Channel1}.  However, if $\tilde{\mathbf{Q}}_{i}$ is far from the optimal $\mathbf{Q}_{i}$, the approximation error can be relatively large.  The bargaining result may be inferior to the one in the non-cooperative game.  This is the case with Channel Set II, as we compare the corresponding sum rates in Fig.~\ref{fig:sumrate_Channel2_bargain} and Fig.~\ref{fig:noncoop_sumrate_Channel2}.

\begin{figure}[!t]
\centering
\includegraphics[width=0.48\textwidth]{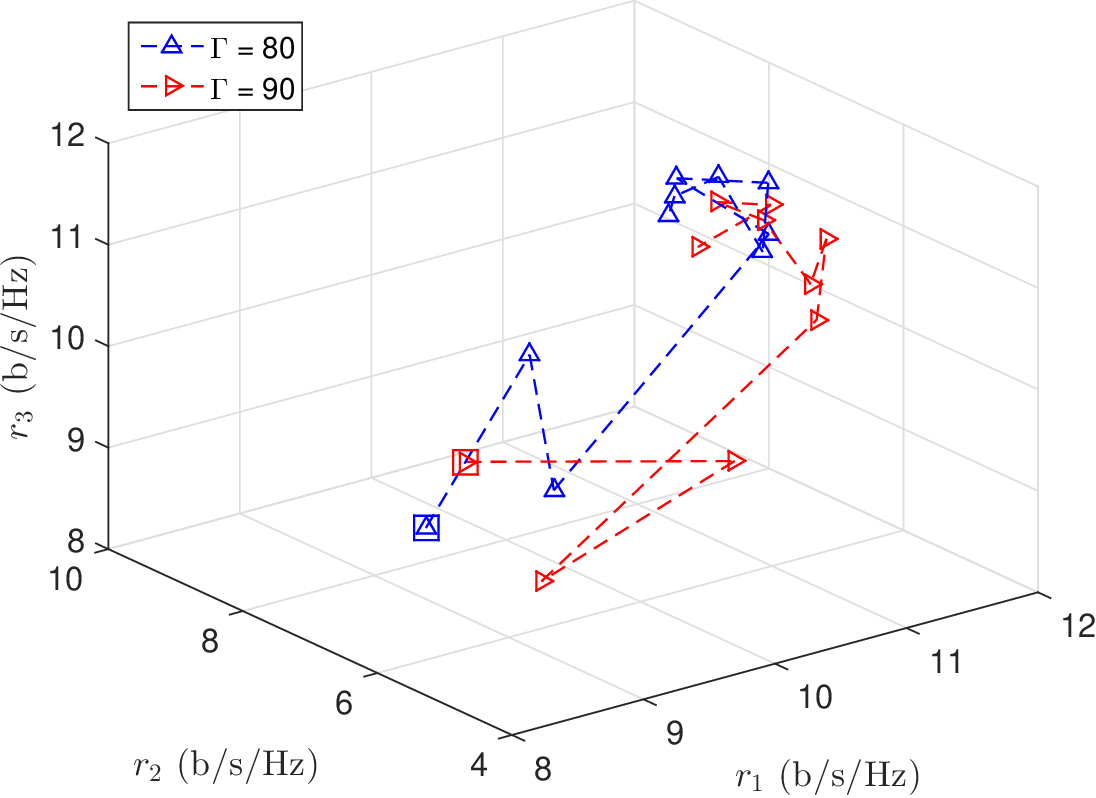}
\caption{Achievable information rates of $K=3$ wireless users in a cooperative game over iterations of updating $\{\tilde{\mathbf{Q}}_{i}\}$.  The starting points are indicated with $\square$.  Channel Set I.}
\label{fig:rate3D_Channel1_bargain_updateQ}
\end{figure}
\begin{figure}[!t]
\centering
\includegraphics[width=0.48\textwidth]{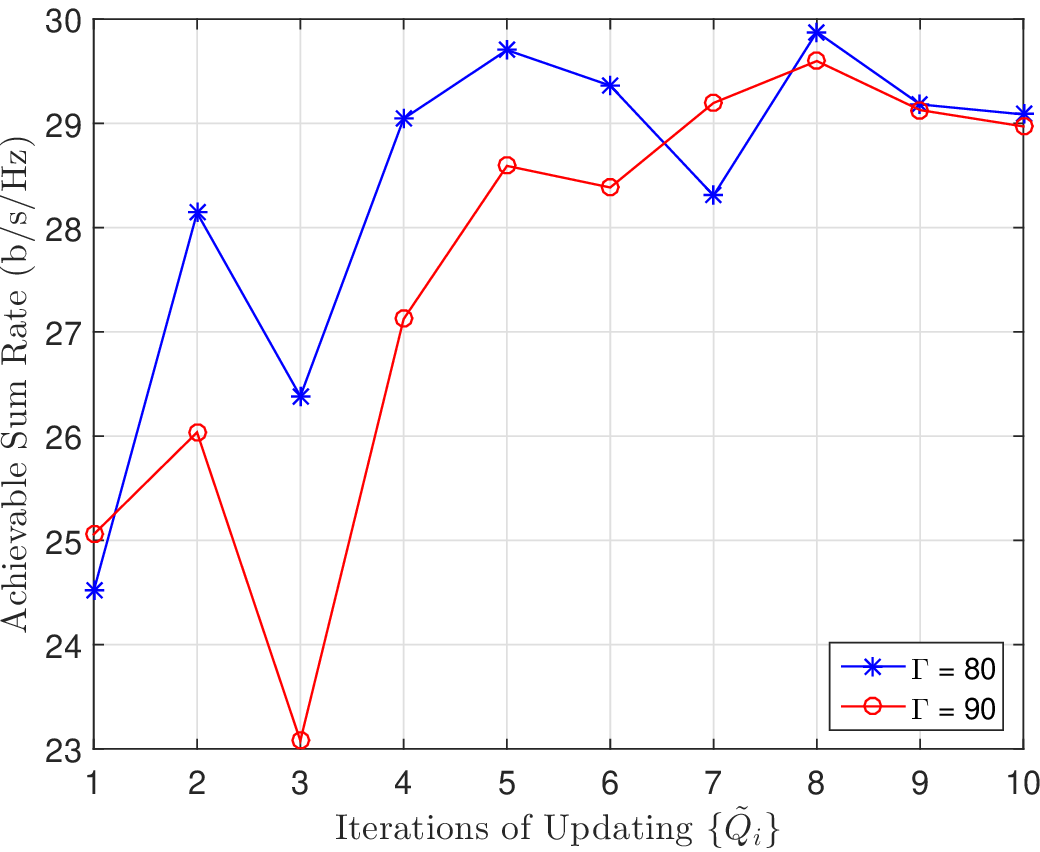}
\caption{Achievable sum rate of $K=3$ wireless users in a cooperative game over iterations of updating $\{\tilde{\mathbf{Q}}_{i}\}$.  Channel Set I.}
\label{fig:sumrate_Channel1_bargain_updateQ}
\end{figure}
\begin{figure}[!t]
\centering
\includegraphics[width=0.48\textwidth]{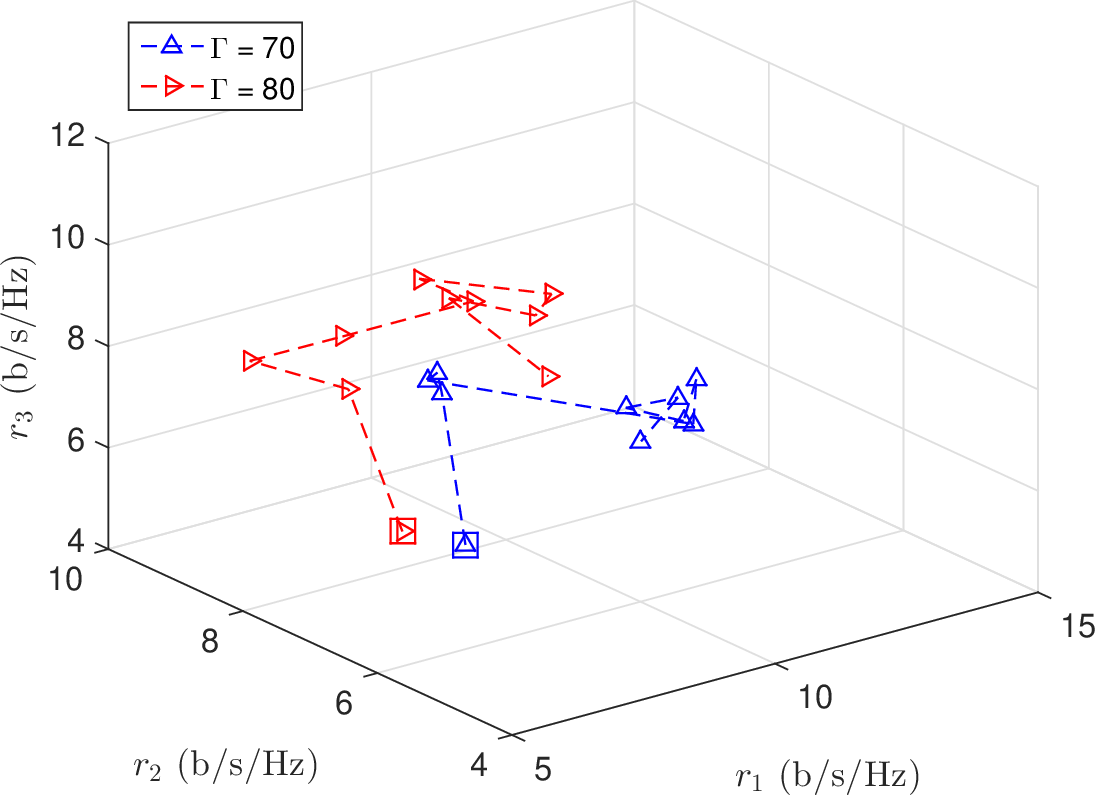}
\caption{Achievable information rates of $K=3$ wireless users in a cooperative game over iterations of updating $\{\tilde{\mathbf{Q}}_{i}\}$.  The starting points are indicated with $\square$.  Channel Set II.}
\label{fig:rate3D_Channel2_bargain_updateQ}
\end{figure}
\begin{figure}[!t]
\centering
\includegraphics[width=0.48\textwidth]{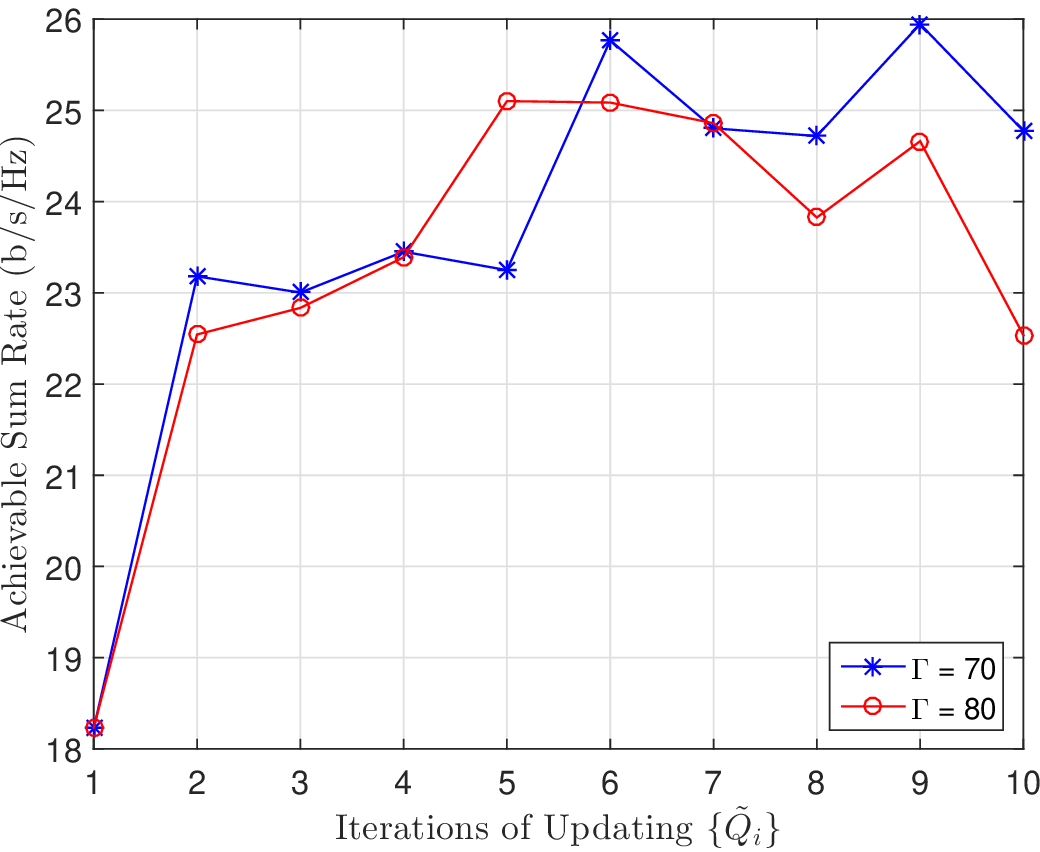}
\caption{Achievable sum rate of $K=3$ wireless users in a cooperative game over iterations of updating $\{\tilde{\mathbf{Q}}_{i}\}$.  Channel Set II.}
\label{fig:sumrate_Channel2_bargain_updateQ}
\end{figure}

To deal with the issue of utility approximation error, $\tilde{\mathbf{Q}}_{i}$ is updated with the previously derived optimal $\mathbf{Q}_{i}$, and the network repeats the distributed bargaining process.  With Channel Set I, Fig.~\ref{fig:rate3D_Channel1_bargain_updateQ} and Fig.~\ref{fig:sumrate_Channel1_bargain_updateQ} show the achievable information rates and the sum rate, respectively, of the three users in the cooperative game over iterations of updating $\{\tilde{\mathbf{Q}}_{i}\}$.  Each point in the figures is a bargaining result.  With Channel Set II, Fig.~\ref{fig:rate3D_Channel2_bargain_updateQ} and Fig.~\ref{fig:sumrate_Channel2_bargain_updateQ} show the achievable information rates and the sum rate, respectively, of the three users in the cooperative game over iterations of updating $\{\tilde{\mathbf{Q}}_{i}\}$.    As illustrated in Fig.~\ref{fig:rate3D_Channel1_bargain_updateQ} and Fig.~\ref{fig:rate3D_Channel2_bargain_updateQ}, the user achievable information rates quickly improve and approach the Pareto front.  Comparing Fig.~\ref{fig:sumrate_Channel1_bargain_updateQ} with Fig.~\ref{fig:noncoop_sumrate_Channel1} and comparing Fig.~\ref{fig:sumrate_Channel2_bargain_updateQ} with Fig.~\ref{fig:noncoop_sumrate_Channel2}, we notice significant improvement in the achievable sum rates resulting from the distributed implementation of the cooperative game.

\section{Conclusions}
\label{sec:conclusion}

In this paper, we have considered the multi-user transmission problem in MIMO interference channels with transmit power constraints and energy-harvesting requirements.  Under mutual interference, each wireless user designs its transmit covariance matrix to maximize the achievable information rate.  At the same time, the transmissions provide sufficient energy for the RF energy harvesters.  The problem is formulated as a strategic game played by the multiple transmitters.  Measuring the interference from others, a transmitter has its best response in the non-cooperative game that constructs the transmit covariance matrix with the multi-level water-filling solution.  An algorithm for the multi-level water-filling is derived.  A pure-strategy Nash equilibrium of the non-cooperative game exists.  It may not be unique or Pareto-efficient.  To find the point of rates that approaches the Pareto front, the users cooperate and bargain over the unit-reward of energy contribution.  The individual utility function is from an approximation of the achievable information rate such that the sum-rate optimization problem can be decomposed and the bargaining process implemented distributively.   To overcome the issue of large approximation error, the bargaining process can repeat with reinitialization using the previously derived optimal solution of the transmit covariance matrices.  With moderate signaling among the users, the cooperative game process quickly reaches a point of rates that is close to the Pareto front.

\appendices

\section{The optimal solution of Problem $\mathcal{P}_{2}$}
\label{ap:Problem2}

To find the optimal solution of Problem $\mathcal{P}_{2}$, we use
the Karush-Kuhn-Tucker (KKT) conditions~\cite{ConvexOpt_Boyd} that are given as ($\forall m$)
\begin{eqnarray*}
\hat{q}_{m} & \geq & 0 \\
\lambda_{m} &\geq & 0 \\
\hat{q}_{m} \lambda_{m} &=& 0 
\end{eqnarray*}
\begin{eqnarray*}
\sum_{m=1}^{M} \hat{q}_{m} - P & \leq & 0 \\
\Gamma - \sum_{m=1}^{M} |\hat{g}_{m}|^{2} \hat{q}_{m} &\leq& 0 \\
\nu_{1} &\geq& 0 \\
\nu_{2} &\geq& 0 \\
\nu_{1}\left( \sum_{i=1}^{M} \hat{q}_{m} - P \right) &=& 0 \\
\nu_{2} \left( \Gamma - \sum_{m=1}^{M} |\hat{g}_{m}|^{2} \hat{q}_{m} \right) &=& 0 \\
-\frac{|\sigma_{m}|^{2}}{1+ |\sigma_{m}|^{2} \hat{q}_{m}} - \lambda_{m} + \nu_{1} - \nu_{2}|\hat{g}_{m}|^{2} &=& 0
\end{eqnarray*}
As $\lambda_{m}$ acts as a slack variable, it can be eliminated that leaves
\begin{eqnarray*}
\hat{q}_{m} & \geq & 0 \\
\hat{q}_{m}\left( \nu_{1} - \nu_{2} |\hat{g}_{m}|^{2} - \frac{|\sigma_{m}|^{2}}{1 + |\sigma_{m}|^{2} \hat{q}_{m}} \right) &=& 0 \\
\nu_{1} - \nu_{2} |\hat{g}_{m}|^{2} - \frac{|\sigma_{m}|^{2}}{1 + |\sigma_{m}|^{2} \hat{q}_{m}} & \geq & 0
\end{eqnarray*}
From these conditions, if follows that $\hat{q}_{m} \geq 1/(\nu_{1} - \nu_{2}|\hat{g}_{m}|^{2}) - 1/|\sigma_{m}|^{2} $.   Therefore, we have
\begin{equation}
\hat{q}_{m}^{*} = \left\{ \begin{array}{l}
  1/(\nu_{1} - \nu_{2}|\hat{g}_{m}|^{2}) - 1/|\sigma_{m}|^{2} , \\ 
   \text{~~~~~~~if~} 1/(\nu_{1} - \nu_{2}|\hat{g}_{m}|^{2}) > 1/|\sigma_{m}|^{2} \\
  0, \text{~~~~if~} 1/(\nu_{1} - \nu_{2}|\hat{g}_{m}|^{2}) \leq 1/|\sigma_{m}|^{2}
\end{array}
 \right.
\end{equation}
where $\nu_{1}$ and $\nu_{2}$ are non-negative parameters that make $\{\hat{q}_{m}^{*}\}$ satisfy the transmit power constraint and the energy harvesting requirement.  This is a water-filing solution with multiple water levels.  For $1/(\nu_{1} - \nu_{2}|\hat{g}_{m}|^{2}) > 1/|\sigma_{m}|^{2}$, the water level is $1/(\nu_{1} - \nu_{2}|\hat{g}_{m}|^{2})$ which depends on $\hat{g}_{m}$.

\section{Discussion on the Estimation of Vector Channel $\mathbf{g}_{i}$}
\label{ap:estimate_g}

As revealed in the optimization problem $\mathcal{P}_{2}$ and in the multi-level water-filling solution \eqref{eq:multilevel-waterfilling}, it is suffice to estimate $\{|\hat{g}_{im}|^{2}\}_{m = 1}^{M_{t}}$ of the vector channel to the RF energy harvester.   The user channel $\mathbf{H}_{ii}$ is estimated and the multi-user interference and noise $\mathbf{R}_{i}$ is measured.  Therefore, the pre-coding matrix $\mathbf{V}_{i}$ can be calculated.   

We need a special arrangement for transmitter $i$ to acquire $\{|\hat{g}_{im}|^{2}\}_{m = 1}^{M_{t}}$.  In this arrangement, only transmitter $i$ transmits, and it concentrates its transmit power $P$ on the $m$th transmission direction represented by the $m$th column of the pre-coding matrix $\mathbf{V}_{i}$. That is, the signal covariance matrix is chosen as $\hat{\mathbf{Q}}_{i} = P\cdot \text{diag}(\underbrace{0,\ldots,0}_{m-1},1,\underbrace{0,\ldots,0}_{M_{t}-m})$ and, accordingly, the transmit covariance matrix is $\mathbf{V}_{i} \hat{\mathbf{Q}}_{i} \mathbf{V}^{H}_{i}$.   The received power at the RF energy harvester is $\zeta \hat{\mathbf{g}}_{i}^{H} \hat{\mathbf{Q}}_{i} \hat{\mathbf{g}}_{i} = \zeta P |\hat{g}_{im}|^{2}$.  Because the RF energy harvester is attached to an IoT device, this value can be fed back to transmitter $i$ through the Internet.

Transmitter $i$ concentrates its transmit power $P$ consecutively on each of the $M_{t}$ transmission directions according to the columns of the pre-coding matrix $\mathbf{V}_{i}$.  Suppose that the transmitter and the RF energy harvester IoT can be synced and the feedback delay can be considered.  The channel elements $\{|\hat{g}_{im}|^{2}\}_{m = 1}^{M_{t}}$ can be estimated and be used in solving Problem $\mathcal{P}_{2}$.

\section{Derivation of the Approximation of the Maximum Information Rate}
\label{ap:TaylorAppro}

The maximum information rate is
\begin{equation}
r_{i} = \log \left| \mathbf{I} + \mathbf{R}_{i}^{-1} \mathbf{H}_{ii} \mathbf{Q}_{i} \mathbf{H}_{ii}^{H} \right|.
\end{equation}
Here, $r_{i}$ is a scalar function of matrix $\mathbf{R}_{i}$.  Therefore, $dr_{i}$ can be given by
\begin{eqnarray}
d r_{i} &=& d  \log \left| \mathbf{I} + \mathbf{R}_{i}^{-1} \mathbf{H}_{ii} \mathbf{Q}_{i} \mathbf{H}_{ii}^{H} \right| \\
&=& \textrm{Tr} \left( (\mathbf{I} + \mathbf{R}_{i}^{-1} \mathbf{H}_{ii} \mathbf{Q}_{i} \mathbf{H}_{ii}^{H} )^{-1} \right. \nonumber \\  
&& \left. d  (\mathbf{I} + \mathbf{R}_{i}^{-1} \mathbf{H}_{ii} \mathbf{Q}_{i} \mathbf{H}_{ii}^{H} ) \right) \label{eq:derivative_log} \\
&=& \textrm{Tr} \left( (\mathbf{I} + \mathbf{R}_{i}^{-1} \mathbf{H}_{ii} \mathbf{Q}_{i} \mathbf{H}_{ii}^{H} )^{-1}  (d \mathbf{R}_{i}^{-1}) \right.  \nonumber \\
&& \left. \mathbf{H}_{ii} \mathbf{Q}_{i} \mathbf{H}_{ii}^{H}  \right) \label{eq:dAZB}\\
&=& - \textrm{Tr} \left( (\mathbf{I} + \mathbf{R}_{i}^{-1} \mathbf{H}_{ii} \mathbf{Q}_{i} \mathbf{H}_{ii}^{H} )^{-1} \mathbf{R}_{i}^{-1} (d \mathbf{R}_{i}) \mathbf{R}_{i}^{-1} \right. \nonumber \\ 
&& \left. \mathbf{H}_{ii} \mathbf{Q}_{i} \mathbf{H}_{ii}^{H}  \right) \label{eq:dZneqone}\\
&=& - \mathrm{Tr} \left(\mathbf{R}_{i}^{-1} \mathbf{H}_{ii} \mathbf{Q}_{i} \mathbf{H}_{ii}^{H} (\mathbf{I} + \mathbf{R}_{i}^{-1} \mathbf{H}_{ii} \mathbf{Q}_{i} \mathbf{H}_{ii}^{H} )^{-1} \right. \nonumber \\ 
&& \left. \mathbf{R}_{i}^{-1} d \mathbf{R}_{i} \right) \\
&=& -\mathrm{Tr} \left( ( \mathbf{R}_{i}^{-1} - (\mathbf{I} + \mathbf{R}_{i}^{-1} \mathbf{H}_{ii} \mathbf{Q}_{i} \mathbf{H}_{ii}^{H} )^{-1} \right. \nonumber \\  
&& \left. \mathbf{R}_{i}^{-1} ) d \mathbf{R}_{i} \right) \\
&=& -\mathrm{Tr} \left( ( \mathbf{R}_{i}^{-1} - (\mathbf{R}_{i} + \mathbf{H}_{ii} \mathbf{Q}_{i} \mathbf{H}_{ii}^{H} )^{-1} ) d \mathbf{R}_{i} \right)
\end{eqnarray}
where, \eqref{eq:derivative_log} follows the fact that $d \log|\mathbf{Z}| = \mathrm{Tr} (\mathbf{Z}^{-1} d\mathbf{Z})$, \eqref{eq:dAZB} follows the fact that $d(\mathbf{AZB}) = \mathbf{A} (d\mathbf{Z}) \mathbf{B}$, and \eqref{eq:dZneqone} follows the fact that $d \mathbf{Z}^{-1} = - \mathbf{Z}^{-1} (d\mathbf{Z}) \mathbf{Z}^{-1}$.  Here, $\mathbf{Z}$ is a complex-valued matrix, and $\mathbf{A}$ and $\mathbf{B}$ are independent of $\mathbf{Z}$ and $\mathbf{Z}^{*}$~\cite{hjorungnes2011complex}.

The first-order Taylor expansion of the maximum information rate $r_{i}$ around $\mathbf{R}_{i} = \tilde{\mathbf{R}}_{i}$ can be written as
\begin{eqnarray}
r_{i} &\approx& r_{i} \vert_{\mathbf{R}_{i} = \tilde{\mathbf{R}}_{i}} + \mathrm{Tr} \left( \nabla_{\mathbf{R}_{i}} r_{i}  \bigg\rvert_{\mathbf{R}_{i} = \tilde{\mathbf{R}}_{i}} \cdot (\mathbf{R}_{i} - \tilde{\mathbf{R}}_{i}) \right) \\
&=& \log \left| \mathbf{I} + \tilde{\mathbf{R}}_{i}^{-1} \mathbf{H}_{ii} \mathbf{Q}_{i} \mathbf{H}_{ii}^{H} \right| \nonumber \\
&& - \mathrm{Tr} \left( (\tilde{\mathbf{R}}_{i}^{-1} - (\tilde{\mathbf{R}}_{i} + \mathbf{H}_{ii} \mathbf{Q}_{i}\mathbf{H}_{ii}^{H} )^{-1}) (\mathbf{R}_{i} - \tilde{\mathbf{R}}_{i}) \right) \nonumber \\
\end{eqnarray}
Replacing $\tilde{\mathbf{R}}_{i}^{-1}  -(\tilde{\mathbf{R}}_{i} + \mathbf{H}_{ii} \mathbf{Q}_{i}\mathbf{H}_{ii}^{H} )^{-1}$ with its approximation $\mathbf{A}_{i} = \tilde{\mathbf{R}}_{i}^{-1}  -(\tilde{\mathbf{R}}_{i} + \mathbf{H}_{ii} \tilde{\mathbf{Q}}_{i}\mathbf{H}_{ii}^{H} )^{-1}$, we have
\begin{equation}
r_{i} \approx \log \left| \mathbf{I} + \tilde{\mathbf{R}}_{i}^{-1} \mathbf{H}_{ii} \mathbf{Q}_{i} \mathbf{H}_{ii}^{H} \right| - \mathrm{Tr}(\mathbf{A}_{i} \mathbf{R}_{i}) + \underbrace{\mathrm{Tr}(\mathbf{A}_{i} \tilde{\mathbf{R}}_{i})}_{c}.
\end{equation}

\section{Sub-gradient Method for the Master Dual Problem}
\label{ap:subgradient}

The energy contribution of transmitter $i$ to RF energy harvester $l$ ($l \in \mathcal{L}$) is denoted as $\beta_{li} = \mathbf{g}_{li}^{H} \mathbf{Q}_{i} \mathbf{g}_{li}$.  Let $\mathbf{b}_{i} = [\beta_{1i}, \beta_{2i}, \ldots, \beta_{Li}]^{T}$.  Let $g_{i}(\{\lambda_{l}\})$ be the dual function obtained as the maximum value of the Lagrangian solved in \eqref{eq:local_problem2} for a given set of $\{\lambda_{l}\}_{l \in \mathcal{L}}$.

The sub-gradient for each $g_{i}(\{\lambda_{l}\})$ is given by
\begin{equation}
\mathbf{s}_{i}(\{\lambda_{l}\}) = \mathbf{b}_{i}(\mathbf{Q}^{*}_{i})
\end{equation}
where $\mathbf{Q}^{*}_{i}$ is the optimal solution of \eqref{eq:local_problem2} for given $\{\lambda_{l}\}$.  Transmitter $i$ broadcasts $\mathbf{b}_{i}(\mathbf{Q}^{*}_{i})$ to other players during each iteration of the cooperative game.  The global sub-gradient is then given by
\begin{equation}
\mathbf{s}(\{\lambda_{l}\}) = \sum_{i \in \mathcal{K}} \mathbf{s}_{i}(\{\lambda_{l}\}) - \mathbf{\Gamma} = \sum_{i \in \mathcal{K}}\mathbf{b}_{i}(\mathbf{Q}^{*}_{i}) - \mathbf{\Gamma}
\end{equation}
where $\mathbf{\Gamma} = [\Gamma_{1}, \Gamma_{2}, \ldots, \Gamma_{L}]^{T}$.  The master dual problem can be solved iteratively as
\begin{equation}
\mathbf{\lambda}^{(z+1)} = \mathbf{\lambda}^{(z)} - \alpha_{z} \mathbf{s}(\mathbf{\lambda}^{(z)})
\end{equation}
where $\mathbf{\lambda} = [\lambda_{1}, \lambda_{2}, \ldots, \lambda_{L}]^{T}$ and $\alpha_{z} \geq 0$ is the step size.













\bibliographystyle{IEEEtran}
\bibliography{IEEEabrv,RFEnergyHarvesting}

\end{document}